\documentclass[iop]{emulateapj}
\usepackage{psfig}

\slugcomment{To appear in the Astrophysical Journal}
\shorttitle{The evolutionary properties of NGC~6791}
\shortauthors{Buzzoni et al.}

\begin{document}

\title{Stellar lifetime and ultraviolet properties of the old metal-rich\\
Galactic open cluster NGC~6791:
a pathway to understand\\ 
the UV upturn of elliptical galaxies\altaffilmark{1}}

\author{Alberto Buzzoni\altaffilmark{2}}
\affil{INAF - Osservatorio Astronomico di Bologna, Via Ranzani 1, 40127 Bologna, Italy}

\author{Emanuele Bertone}
\affil{INAOE - Instituto Nacional de Astrof{\'\i}sica {\'O}ptica y Electr{\'o}nica, 
Luis Enrique Erro 1, 72840 Tonantzintla, Puebla, Mexico}

\author{Giovanni Carraro\altaffilmark{3}}
\affil{ESO - European Southern Observatory, Alonso de Cordova 3107, 
Casilla 19001, Santiago 19, Chile}

\and

\author{Lucio Buson}
\affil{INAF - Osservatorio Astronomico di Padova, Vicolo Osservatorio 5,
35122 Padova, Italy}

\altaffiltext{1}{Based on observations carried out at the Italian 
Telescopio Nazionale Galileo, operated by INAF at the Roque de los Muchachos 
Observatory (La Palma, Spain)}
\altaffiltext{2}{e-mail: alberto.buzzoni@oabo.inaf.it}
\altaffiltext{3}{On leave from Dip. di Astronomia, Universit\'a di Padova, Italy}

\begin{abstract}
The evolutionary properties of the old metal-rich Galactic open 
cluster NGC~6791 are assessed, based on deep $UB$ photometry and 2{\sc Mass} 
$JK$ data. For 4739 stars in the cluster, bolometric luminosity and effective 
temperature have been derived from theoretical $(U-B)$ and $(J-K)$ color fitting.
The derived H-R diagram has been matched with
the {\sc Uvblue} grid of synthetic stellar spectra to obtain the integrated SED of 
the system, together with a full set UV (Fanelli) and optical (Lick)
narrow-band indices.

The total bolometric magnitude of NGC~6791 is $M^{\rm bol}_{6791} = -6.29$, 
with a color $(B-V)_{6791} = 0.97$. The cluster appears to be a fairly good proxy of 
standard elliptical galaxies, although with significantly bluer 
infrared colors, a shallower 4000~\AA\ Balmer break, and a lower $Mg_2$ index.
The confirmed presence of a dozen hot stars, along their EHB 
evolution, leads the cluster SED to consistently match the properties of the most 
active UV-upturn galaxies, with 1.7{\scriptsize $\pm 0.4$}\% 
of the total bolometric luminosity emitted shortward of 2500~\AA. 

The cluster Helium abundance results $Y_{\rm 6791} = 0.30${\scriptsize $\pm 0.04$}, 
while the Post-MS implied stellar lifetime from star number counts fairly agrees 
with the theoretical expectations from both the {\sc Padova} and {\sc Basti} stellar 
tracks. A Post-MS fuel consumption of 0.43{\scriptsize$\pm 0.01$}~M$_\odot$ is 
found for NGC~6791 stars, in close agreement with the estimated 
mass of cluster He-rich white dwarfs. Such a tight figure may lead to suspect that a 
fraction of the cluster stellar population does actually not reach the minimum mass 
required to effectively ignite He in the stellar core.
\end{abstract}

\keywords{galaxies: elliptical and lenticular, cD --- galaxies: evolution --- Galaxy: open clusters 
and associations: individual (NGC~6791) --- stars: evolution --- ultraviolet: galaxies
}

\section{Introduction}

As a natural marker of the hot stellar component, the study of the ultraviolet (UV) 
properties of galaxies and smaller star clusters provides us with an important piece of information 
to more deeply constrain the overall evolutionary status of a stellar aggregate as a whole.

While the implied presence of O-B stars hotter than 30\,000 - 40\,000~K is a quite standard 
condition for any young and/or star-forming system, like in spiral and irregular galaxies 
\citep{code82}, things revealed to be much more puzzling when facing the UV-enhanced luminosity 
as sometimes observed in the spectral energy distribution (SED) of old quiescent elliptical galaxies.
Although within a wide range of strength, the abrupt rise in the UV emission 
of ellipticals and spiral bulges shortward of 2000~\AA, known since the 80's as the ``UV upturn'' 
phenomenon \citep{code79,bertola82}, seems to characterize the stellar population
in old metal-rich environments. This poses a crucial constrain to stellar evolution theory
in order to settle this effect into a convenient theoretical framework.

Among the different UV sources one can run into in any old stellar population 
\citep[see, e.g.][for an updated review]{yi04}, hot Post-AGB (PAGB) and extreme horizontal-branch 
(EHB) stars stand out as main contributors to short-wavelength emission. Depending on the 
core/envelope mass ratio ($M_c/M_e$), evolution of low-mass ($M\la 2~M_\odot$) stars actually heads to the 
final white-dwarf (WD) stage either through a full completion of the AGB phase and the subsequent 
planetary-nebula event \citep[hot-PAGB evolution, see e.g.][]{pac70,iben83},
or by skipping partially or {\it in toto} any AGB evolution \citep{greggio90}. EHB stars are the 
natural outcome in the latter case, being on their route toward the WD cooling sequence 
after completing the horizontal-branch (HB) evolution \citep{dorman95,dcruz96}.
In both cases, stellar structure is characterized by a shallow external envelope, which lets the 
hot H-burning shell to ``emerge'' close to the stellar surface thus moving stars toward the 
high-temperature region of the H-R diagram.

In the c-m diagrams of a few globular clusters-- $\omega$~Cen \citep{dcruz00} and NGC~2808 
\citep{harris74,castellani06a} are good examples in this sense-- EHB stars appear to clump at 
the hot-temperature edge of the HB being classified, from the spectroscopic point of view, 
as hot subdwarfs (sdB). 
Spectroscopy \citep{brown97} and imaging \citep{brown00} of resolved c-m diagrams for 
stellar populations in local galaxies, like M32, seem definitely to point to these stars
as the main responsible for the UV upturn providing some (unknown) critical threshold
in $[Fe/H]$ to be exceeded to trigger the phenomenon 
\citep[e.g.][]{greggio90,bressan94,buzzoni95,buzzoni07,dorman95}.

Although metallicity might be the leading parameter in the game, the exact physical mechanisms
at work still lack any firm observational confirmation as photometry of evolved UV-bright 
stars in external galaxies still confine to the relevant case of M31 and its satellite system 
\citep[e.g.][]{bertola95,brown98,brown00}, and no suitable spectroscopy for single 
stars is available to date.
The study of old metal-rich open clusters in the Galaxy may provide important clues in this
regard. Although over a much reduced mass scale, open clusters may prove, in fact, 
to be an effective proxy to constrain the nature of the again-emerging UV in 
elliptical galaxies and spiral bulges. 

In this framework, NGC~6791 plays perhaps a central role being, with its $\sim 4000$~M$_\sun$ 
\citep{kinman65,kaluzny92}, among the most massive and populated open clusters 
in our Galaxy. Its old age, about 8~Gyr, has recently been confirmed by \citet{chaboyer99},
\citet{anthony07} and \citet{grundahl08}, relying on CCD photometry, while extensive high-resolution 
spectroscopy 
of member stars \citep{peterson98,carraro06,origlia06,gratton06,boesgaard09} points to a supersolar 
metallicity (i.e.\ $[Fe/H]\sim +0.4\pm0.1$). Located about 4~kpc away 
\citep{carraro99,carraro06,carney05} in an oscillating orbit, that periodically brings it 
closer to the bulge of the Milky Way \citep{bedin06,wu09}, NGC~6791 stands out as a sort of backyard 
``Rosetta Stone'' to assess the UV emission of more distant ellipticals. 

This match is even more reinforced by the truly peculiar hot-HB content of this cluster, 
with a sizeable fraction of sdB/O stars, as first detected by \citet{kaluzny92} 
and \citet{kaluzny95}.
On the basis of ground and space-borne (UIT and HST) UV observations 
\citep{yong00,liebert94,landsman98}, these hot sources have then been interpreted 
as EHB stars with $T_{\rm eff} \sim 24$-32\,000~K.
Some hints for the presence of a composite stellar population in NGC~6791 recently
came from \citet{twarog11}, who reported a significant color shift in the 
Main-sequence location along the cluster radius. This feature may lead to
an age $\sim 1$~Gyr older in the core region compared to cluster periphery,
although an alternative explanation in terms of reddening gradient may not be 
firmly excluded \citep{platais11}.

Facing the relevance of NGC~6791 as a potentially effective proxy of UV-upturn
ellipticals, in this note we present the results of an application of population 
synthesis techniques to further explore the evolutionary lifetime of the cluster
stellar population and the integrated UV properties of the system as a whole. 
For our task we take advantage of a new set of deep $UB$ CCD photometry, resolving
stars down to $B \sim 22$, complemented at bright luminosity by the 2{\sc Mass} IR 
observations. This observing dataset is briefly described in Sec.~2, referring
the reader to an accompanying paper (Buson et al. 2011, Paper II, in preparation) for more details.
By calibrating the apparent color of our stars with a set of theoretical models
we have been able, in Sec.~3, to convert the c-m diagram of the cluster into 
the H-R fundamental plane of bolometric luminosity and effective temperature.
This allowed us to single out and study in fairly good detail the hot stellar component
of the cluster, and lead to the integrated SED of the whole aggregate, by summing
up the spectrophotometric contribution of each star (Sec.~4). By relying on the synthetic 
broad-band colors and narrow-band UV and optical spectrophotometric indices, we assessed 
in Sec.~5 the problem of how close does NGC~6791 resemble the case of UV-upturn
elliptical galaxies. Finally, a careful match of the star number counts across
the cluster H-R diagram and the theoretical prescriptions of the so-called
``fuel consumption theorem'' of \citet{rb86} led us to derive, in Sec.~6, the implied 
stellar lifetime along the different evolutionary branches of the diagram, together
with an estimate of the corresponding energetic constraints from the consumed 
nuclear fuel inside stars. A number of interesting implications on the distinctive
evolutionary properties of the cluster are therefore discussed in this section, 
and further expanded in the final summary of our conclusions, in Sec.~7.

\begin{figure}
\centerline{
\psfig{file=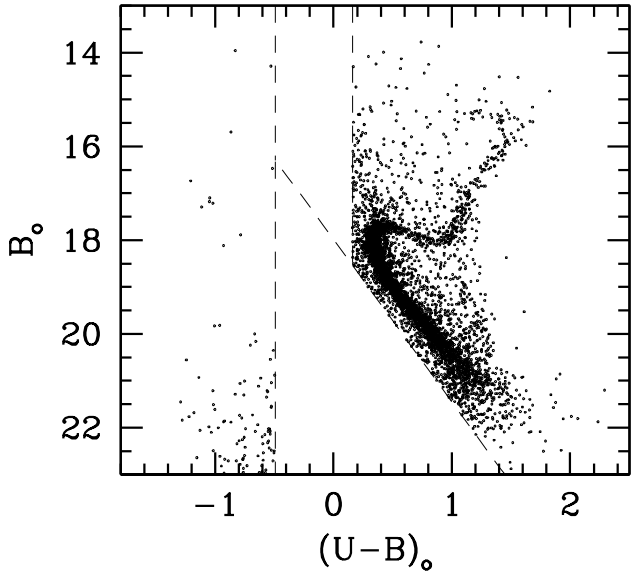,width=\hsize}
} 
\caption{
The observed $B$ vs.\ $(U-B)$ c-m diagram of NGC~6791, according to our deep TNG
photometry. Magnitude and color have been corrected for reddening according to
\citet{twarog09}. To statistically pick up the genuine cluster stellar population
we applied to the plot a selection in the observed (i.e.\ non-dereddened) color domain
by rejecting as possible field interlopers all the stars included in a $-0.4 \la (U-B) \la +0.25$ 
color strip plus those objects lying below the MS locus, as sketched in the diagram.
\label{f1}}
\end{figure}

\section{Observing database and cluster selection}

A set of deep CCD $U,B$ imagery of NGC~6791 has been collected at the 3.6m Telescopio
Nazionale Galileo (TNG) of La Palma (Spain) along the three nights of July 29-31, 2003.
The cluster field has been sampled by stacking four (partially overlapping) 
$9^\prime.4\times 9^\prime.4$ quadrants on the sky taken with the LRS FOSC camera 
equipped with a 2048 $\times$ 2048 back-illuminated E2V CCD. This setup provided a 
platescale of $0^{\prime\prime}.275$px${^-1}$ across a total field of 
$17^\prime.0\times17^\prime.0$ (roughly $20\times 20$~pc across, at the cluster distance). 
Exposure time for the $U$ observations
has been 1200 sec, while $B$ frames have been taken with 300 sec.
A standard photometric reduction of the images has been carried out, relying on 
\citet{landolt92} reference fields, and by matching a supplementary set of 
16 cluster stars in common with the photoelectric observations of \citet{montgomery94}.

Data have been reduced with the {\sc Iraf}\footnote{{\sc Iraf} is distributed by NOAO, which are 
operated by {\sc Aura} under cooperative agreement with the NSF.} packages {\sc Ccdred}, {\sc Daophot}, 
{\sc Allstar} and {\sc Photcal} using the point spread function (PSF) method \citep{stetson87}. 
This led to a photometric catalog with 7831 entries of stars down to $B \sim 22$~mag
with measured $U-B$ color.
For reader's better convenience, a detailed discussion of this database is deferred to 
Paper II, just focussing here to the application of the photometric catalog to our 
cluster spectral synthesis procedure.

For its low Galactic latitude $(l,b) = (70^o,+10.9^o)$ NGC~6791 is heavily embedded
into Galaxy disk. This is evident at first glance from the resulting c-m diagram of
the sky region surrounding the cluster (see Fig.~\ref{f1}, and 
even more clearly \citealp{kaluzny95}).
A comparison with synthetic c-m diagrams of the same region, as from the \citet{girardi05} 
Galactic model (see Paper II for details), makes clear that a major contamination 
along the cluster sightline comes from bright
A-F stars belonging to the thick-disk stellar component, while a negligible
contribution from stars of later spectral type has to be expected, especially as 
foreground interlopers. This provides us with a simple diagnostic scheme as one
can imagine to (mostly) pick up the {\it bona fide} cluster stellar population 
by rejecting field A-F stars in a $-0.4 \la (U-B) \la +0.25$ color strip
\citep[see, e.g.][for the relevant fiducial colors]{johnson66} and all ``redder'' 
(i.e. later type) objects fainter than the fiducial main sequence (MS) locus 
of the cluster, as sketched in Fig.~\ref{f1}.

With these magnitude/color cuts, we are eventually left with a total of 5202 stars
in our $UB$ catalog.
After reddening correction, assuming a color excess $E(B-V) = 0.125$ 
or $E(U-B) = 0.09$ from \citet{twarog09} this {\it bona fide} star sample 
provides an integrated magnitude and color of $[B_o, (U-B)_o] = [8.77, 0.56]$. 

Being especially addressed to the study of the hot stellar component in the cluster,
our $UB$ imagery is of course not well suited to match also the cool red giant stars.
To a more accurate check of the resulting c-m diagram, in fact, most of these stars 
suffered from poor photometry as their exceedingly red colors often required some prudent
extrapolation of our photometric solution. In addition, the reddest stars were sometimes barely 
detectable in the $U$ frames or conversely resulted partly saturated in $B$. For this reason, 
the bright stellar sample in the cluster field (including basically all the RGB+AGB stars 
for $B \ga 15$) have been directly assessed through the infrared observations of the 2{\sc Mass} 
survey \citep{2mass}. 

A selection of all the 2{\sc Mass} stellar candidates across our field with 
color $(J-K) \ge 0.35$ (that is consistent with or redder than the cluster Turn Off point)
provided 94 stars, of which 61 objects were also catched by our $UB$ survey, although in a few 
cases (3 stars) as saturated objects.
The remainig 33 objects were actually not comprised in our $UB$ catalog being all 
bright stars sampling the reddest tail in the overall color distribution.

\begin{figure}
\centerline{
\psfig{file=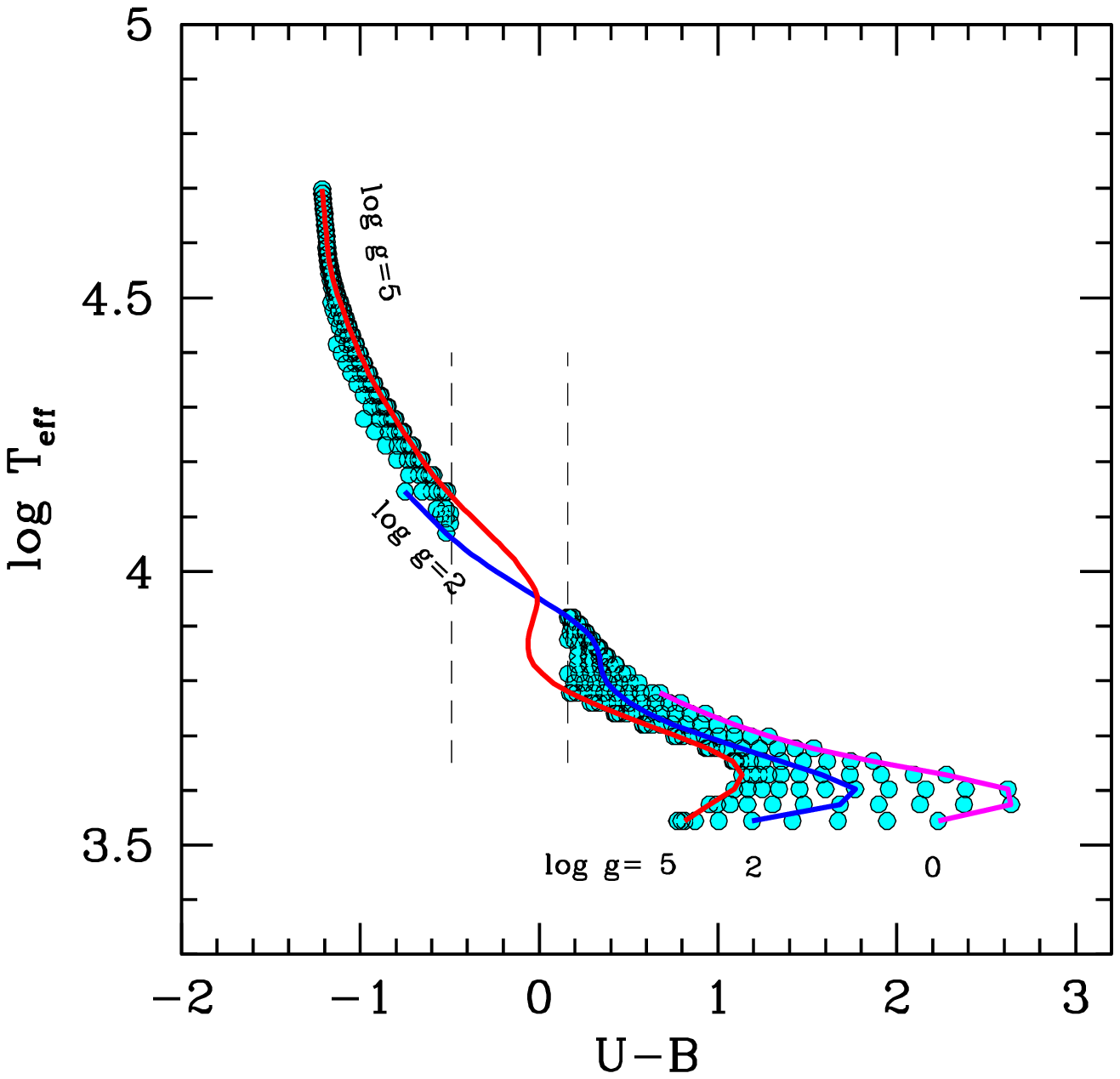,width=0.9\hsize}
} 
\centerline{
\psfig{file=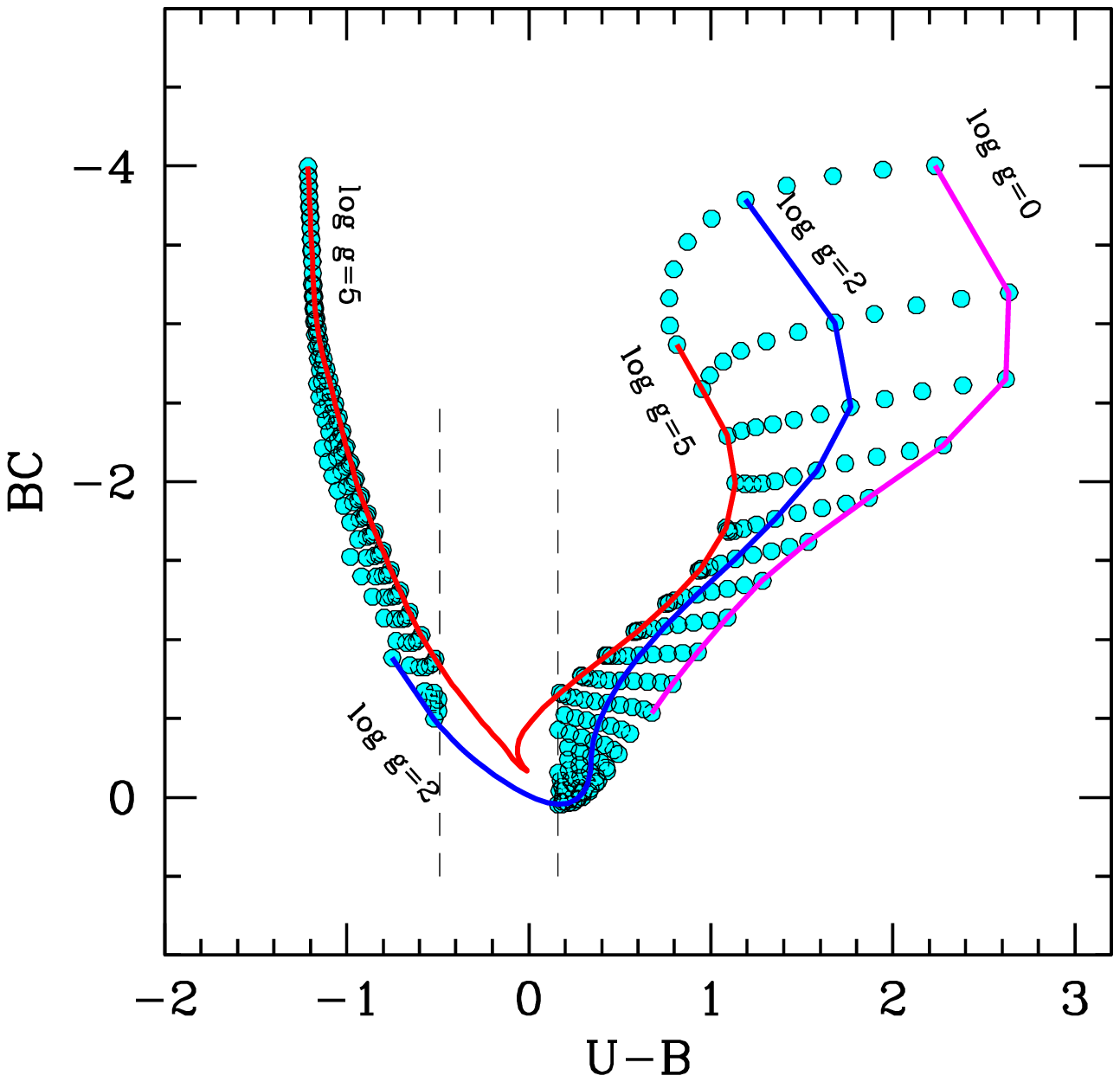,width=0.9\hsize}
} 
\caption{
Theoretical relationship between $(U-B)$ color, effective temperature {\it (upper panel)} 
and $B$-band bolometric correction {\it (lower panel)} for stars with $[Fe/H] = +0.4$
according to the {\sc Uvblue} grid of synthetic stellar spectra \citep{rodriguez05}
in its updated version with the \citet{castelli03} revised chemical opacities.
The model sequences for fixed values of surface gravity, namely $\log g = 0, 2$ (giants), 
and 5 (dwarfs) are singled out and labelled on the plots for reader's better reference.
The assumed ``zone of avoidance'' for field interlopers has been marked in each grid.
\label{f2}}
\end{figure}

\section{The fundamental parameters of cluster stars}

In order to lead to a self-consistent spectral synthesis of the cluster stellar
population, apparent magnitudes and colors of member stars have to be converted
to the fundamental physical plane of the H-R diagram such as to safely derive for
each star its intrinsic parameters (namely bolometric luminosity,
effective temperature and surface gravity).
An iterative algorithm has been set up in this regard, with the help of the
{\sc Uvblue} synthetic library of stellar spectra \citep{rodriguez05}, 
computed for $[Fe/H] = +0.4$ and with the \citet{castelli03} revised chemical 
opacities, as shown in Fig.~\ref{f2}.

\begin{figure*}
\centerline{
\psfig{file=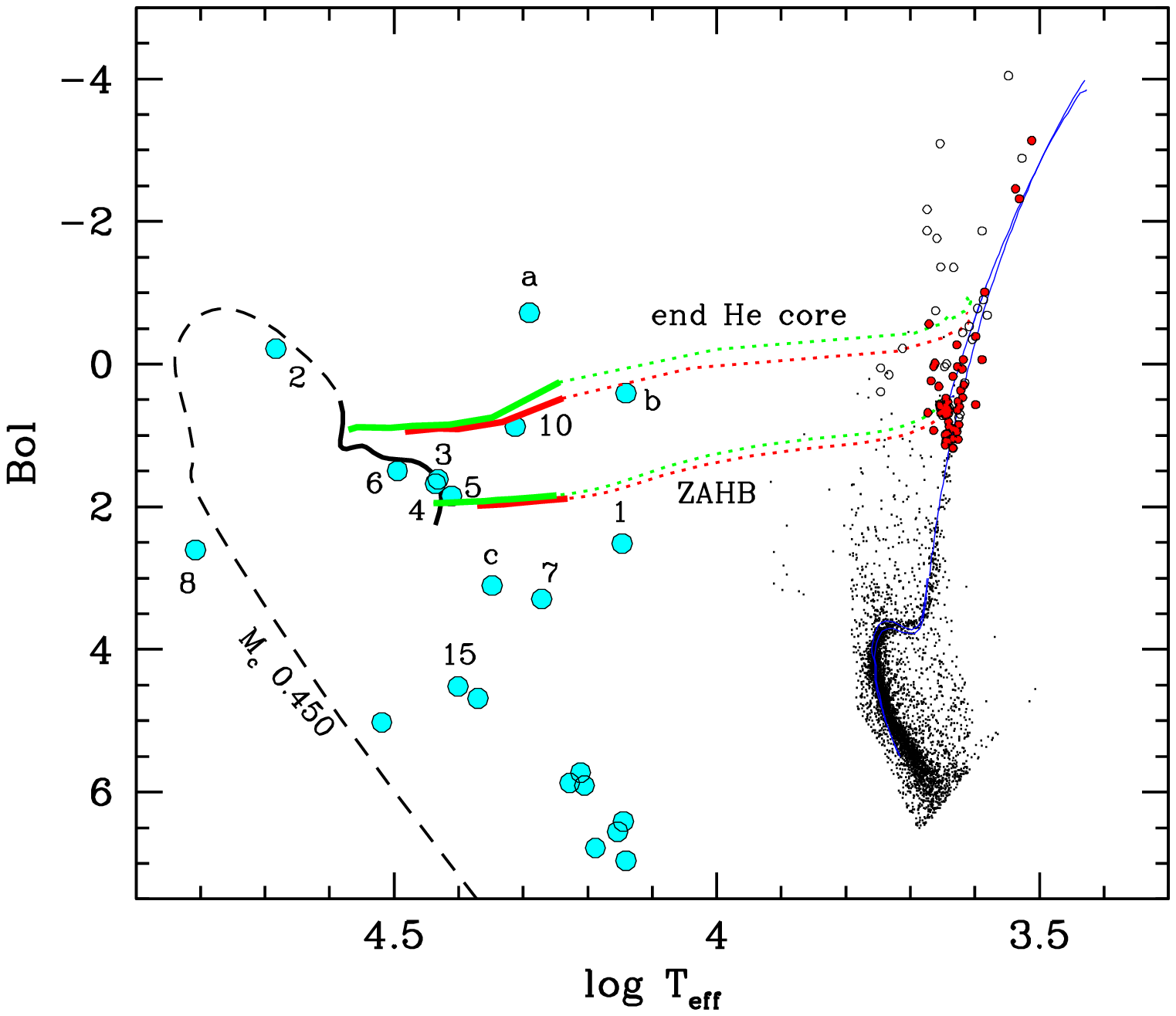,width=0.56\hsize}
} 
\caption{
The derived H-R diagram of NGC~6791. Big (blue) dots mark the hot stellar component with
$T_{\rm eff} \ga 10,000$~K. When available, stars are labelled with their ID number from the
\citet{kaluzny92} catalog. Three new stars, consistent with hot-HB evolution, appear
in our survey and are labelled with ``a'', ``b'', and ``c'' in the plot.
Small dots in the red-giant region of the diagram indicate the 94 2{\sc Mass} stars that integrated
our $UB$ photometry, as discussed in the text. Of these, the 61 stars in common with the $UB$
sample appear as (red) solid dots. The theoretical isochrones from the
{\sc Padova} database \citep{bertelli08} for ($Z,Y$) = (0.04, 0.34) and $t = 6$ and 8~Gyr
are overplotted to the data, together with the expected HB evolutionary strip between  
the ZAHB locus as a lower edge and the He core exhaustion locus as an upper envelope.
Calculations for HB models are from the {\sc Basti} database \citep{pietrinferni07}
for a ($Z,Y$) = (0.04, 0.30) chemical mix with solar-scaled \citep[red curve, from][]{pietrinferni04}
and $\alpha$-enhanced \cite[green curve, from][]{pietrinferni06} metal partition. 
The case of a $0.45~M_\odot$ star evolving as an EHB (and AGB-{\it manqu\'e}) object is also 
displayed in some detail (thick black solid line), according to \citet{dcruz96} for $[Fe/H] = +0.37$.
See text for a discussion.
\label{f3}}
\end{figure*}

To avoid spurious transformations in consequence of high color uncertainty 
of faint objects we restrained our analysis only to stars in our {\it bona fide} sample 
brighter than $B = 21.5$, namely about 3.5~mag below the Turn Off point.
These are 4706 entries in total, from the $UB$ catalog, of which 61 objects overlap the 
2{\sc Mass} sample. For each star we proceeded first by fixing a 
reference value for surface gravity, namely $\log g = 3$. 
With this value, and the (dereddened) $(U-B)$ color of the star we then entered the two
panels of Fig.~\ref{f2} deriving a guess value for $\log T_{\rm eff}$ and the $B$-band 
bolometric correction, $BC = Bol-B$. With the help of $BC$, and once accounting for the 
reddening and distance modulus, the latter being $(m-M) = 13.07$ as from \citet{twarog09},
the $B$ apparent magnitude can eventually be converted
to $\log L/L_\odot$. At this point, a self-consistency check must hold as, by definition,
\begin{equation}
\log (L/L_\odot) = 2\,\log (R/R_\odot) +4\,\log (T_{\rm eff}/T_\odot)
\label{eq:1}
\end{equation}
As the stellar radius $R$ is tied to $g$ through $g= GM/R^2$, then eq.~(\ref{eq:1}) 
can easily be re-written in terms of stellar surface gravity as
\begin{equation}
\log (L/L_\odot) = \log (M/M_\odot) -\log (g/g_\odot) +4\log (T_{\rm eff}/T_\odot),
\label{eq:2}
\end{equation}
where we assume $\log g_\odot = 4.45$ in c.g.s. units and $T_\odot = 5780$~K for the Sun.
Note that eq.~(\ref{eq:2}) only weakly depends on the exact value
of the stellar mass ($M$). For example, even a factor of two uncertainty on $M$ 
would only reflect into a $\Delta \log g \sim \pm 0.3$ uncertainty, which is safely
within the sampling step of the {\sc Uvblue} model grid (i.e. $\Delta \log g = 0.5$~dex).
Considering the estimated age for NGC~6791 and the implied Turn Off mass, then we 
could conservatively set $M = 1$~M$_\odot$ with no relevant impact on our conclusions.

At this point, by entering eq.~(\ref{eq:2}) with the guess values for $\log T_{\rm eff}$ and
$\log L$, we get back a refined (``predicted'') estimate 
for $\log g$ to be used for a new iteration, such as 
\begin{equation}
\log g_{\rm pred} = 4\log T_{\rm eff} -\log (L/L_\odot) -10.60.
\label{eq:gg}
\end{equation}

The iterative procedure has been carried on for each entry of our {\it bona fide}
catalog until the ``predicted'' $\log g$ matches, within $\pm 0.25$~dex, the previous guess value.
Accordingly, this also left us with the final values for $\log L$ and $\log T_{\rm eff}$.
It could be verified that a solution is quickly achieved in a couple of iterations or so,
providing the star locates into the sampled domain of {\sc Uvblue} models.
This has actually always be the case, except for the little bunch of WD stars 
easily recognized in the c-m diagram of Fig.~\ref{f1} fainter than $B \ga +20$
and with negative $(U-B)$ colors. Their gravity is far larger ($\log g \sim 6$-7)
than the {\sc Uvblue} domain so that they have forcedly to be approximated 
within the theoretical grid with $\log g = 5$ models.

A different procedure has been applied, on the contrary, for the 94 stars included in the
2{\sc Mass} sample. For them we relied on the accurate bolometric
calibration by \citet{buzzoni10}, who accounted for NGC~6791 red giants themselves
in their analysis. According to the (dereddened) $(J-K)$ color, we obtain for our cluster
the following transformation equations:\footnote{In addition to the luminosity and temperature
estimates as from eq.~(\ref{eq:lt}), also a measure of surface gravity for these stars can
be obtained, via eq.~(\ref{eq:gg}).}

\begin{equation}
\left\{
\begin{array}{l}
\log T_{\rm eff} = -0.248\, (J-K) +3.82 \\
\hfill \left[ \rho,\sigma \right] = \left[ 0.97, 0.012\,{\rm dex} \right]\\
~~ \\
(Bol-K) = 1.594\, (J-K)+1.13\\
\hfill \left[ \rho,\sigma \right] = \left[ 0.98, 0.04\,{\rm mag} \right]\\
\end{array}
\right.
\label{eq:lt}
\end{equation}

A comparison of the two calibration procedures, for our 58 2{\sc Mass} stars with available 
$(J-K)$ and unsaturated $(U-B)$ color confirmed that effective temperature is 
derived within an internal accuracy of $\sim$6\% (namely 
$\langle \Delta \log T_{\rm eff}\rangle = +0.011 \pm 0.027$~dex,
in the sense ``$UB$ - $JK$''), while the bolometric magnitude scale is
reproduced within a $\langle \Delta Bol \rangle = -0.12 \pm 0.22$~mag accuracy.

In total, bolometric luminosity and effective temperature have been secured for 
4739 stars, whose 4645 entries come from the $UB$ {\it bona fide} sample and 94 stars from the 
2{\sc Mass} database.\footnote{Note that for the 61 stars with available $UB$ and $JK$ photometry,
the latter calibration was eventually retained for our analysis.}

\subsection{The cluster H-R diagram}

The resulting H-R diagram from our iterative conversion of the $B$ vs. 
$(U-B)$ plot of  Fig.~\ref{f1} plus the bright 2{\sc Mass} extension
is displayed in Fig.~\ref{f3}.
As a reference guideline, we superposed to the stellar distribution 
two isochrones from the Padova database \citep{bertelli08}, for 
$t = 6,$ and 8~Gyr and $(Z, Y) = (0.04, 0.34)$, closely matching the cluster metallicity. 
The fairly good agreement confirms the reliability of our transformation procedure.
The Turn Off point appears to have an effective temperature 
$T_{\rm eff} \simeq 5700${\scriptsize $\pm 50$}~K, consistent with a spectral type G5 \citep{johnson66}. 

Just a glance to the figure makes evident one outstanding feature of NGC~6791
stellar distribution, with a marked deficiency of bright red giants, surmounting the 
luminosity level of the red HB clump.
This scanty population of bright cool stars (assumed to include both the bright tail of
RGB evolution and the full AGB phase), compared for instance with the HB population, is 
in fact an established feature of NGC~6791 \citep{kalirai07}, widely recognized also in 
other observing sets taken in different photometric bands
\citep[see, for instance, the c-m diagram of][]{montgomery94,kaluzny95,stetson03,platais11}.
These arguments led \citet{kalirai07} to envisage a sort of physical link, essentially
driven by enhanced stellar mass loss, between the peculiar RGB morphology of NGC~6791 
and a prevailing presence of low-mass ($\langle m_{\rm WD}\rangle = 0.43 \pm 0.06$~M$_\odot$)
He-rich WDs, a feature that might call for an anticipated completion
of RGB evolution, missing in some cases even the canonical He-flash event.

Although within the large uncertainties due to the low statistics and the intervening
effect of field contamination, we think that even a raw estimate of the HB vs.\ bright
RGB+AGB relative number counts can offer an interesting alternative interpretation being
the ratio strongly modulated by the Helium abundance of the cluster stellar population. As a well 
established result \citep{iben68}, in fact, stellar evolution theory tends to predict
a quicker RGB evolution at bright luminosity with increasing Helium abundance.
Consequently, a larger $R = N_{\rm HB}/N_{\rm RGB}$ ratio has to be expected in 
Helium-rich stellar systems. 
As we will show in better detail in Sec.~7, this might actually be the case
for NGC~6791.

\subsection{The hot stellar component}

The H-R diagram of Fig.~\ref{f3} allows us a quite clean assessment of the hot
stellar component in the cluster. This recognition is actually favored by the
good sensitivity of the $(U-B)$ color to the temperature range of B stars,
hotter than $T_{\rm eff} \sim 10,000$, as shown in Fig.~\ref{f2}.
In particular, the HB morphology can be accurately traced by our calibration,
although a few possible A-type member stars may have been missed according 
to our color selection, as explained in Sec.~2.

\begin{figure}
\centerline{
\psfig{file=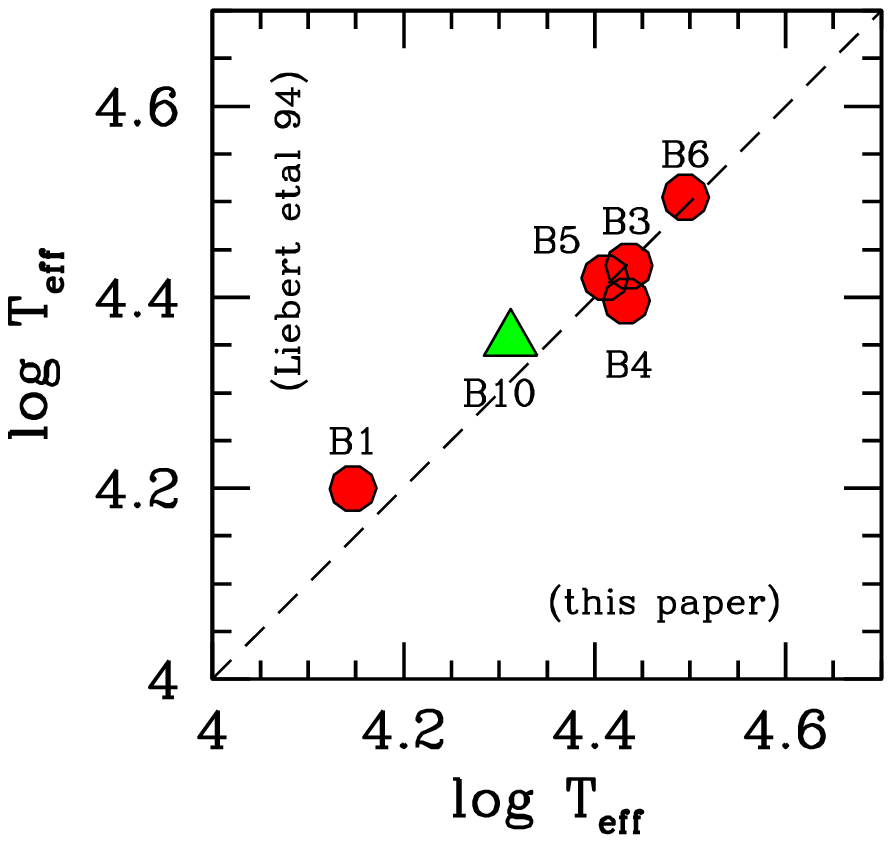,width=0.9\hsize}
} 
\caption{
The \citet{liebert94} temperature scale from high-resolution spectroscopy 
is compared with our $(U-B)$-based calibration for the five hot stars in the 
\citet{kaluzny92} list, as labelled (big dots). One more star in the list 
(B10) has been made available from the \citet{landsman98} spectroscopy and has been 
added to the plot (big triangle).
Spectroscopic temperature estimates are confidently reproduced by our photometric
calibration within a $\pm 7$~\% (1 $\sigma$) relative scatter.
\label{f4}}
\end{figure}

A straight comparison of our data can be performed in Fig.~\ref{f3} with the theoretical 
ZAHB and the upper envelope tracing the end of the core He-burning stage, according to
the {\sc Basti} model database \citep{pietrinferni07}.
A standard solar-scaled chemical composition has been assumed, as from 
\citet{pietrinferni04}, with $(Z, Y) = (0.04, 0.30)$, together with the 
$\alpha$-enhanced case \citep{pietrinferni06}. For both model sets we singled out 
with tick solid lines in the figure the locus of $M \le 0.48$~M$_\odot$ stars which, 
for this metallicity, identifies the EHB objects. 
At least six stars in our sample seem to match this constrain. They are all comprised in
the list of UV-bright stars proposed by \citet{kaluzny92}, with ID code B02, B03, B04, B05,
B06, and B10 (see Paper II for a detailed discussion).
Most of them have been spectroscopically characterized by \citet{liebert94} and 
\citet{landsman98}
leading to an accurate temperature classification. As an important cross-check of our
transformation procedure, we compared in Fig.~\ref{f4} our temperature scale
obtained from the $(U-B)$ color and the corresponding estimate by \citet{liebert94} 
from the spectral fit of the Balmer absorption lines with synthetic high-res spectra. 
The agreement is most than encouraging, with a relative scatter of $\pm 7$~\%
(at 1-$\sigma$ level) in the inferred values of $T_{\rm eff}$ with the two methods.

A closer inspection of Fig.~\ref{f3} reveals, however, that a few other stars 
(three new ones marked with letter ``a'', ``b'', and ``c'' in the plot plus
the Kaluzny object B08) may still be viable hot-HB members owing to the photometric 
uncertainty, that somewhat scatter the point distribution 
in the diagram. Two more objects in this region (namely stars B01 and B07 in 
the \citealp{kaluzny92} original list) have been first claimed to be field stars 
by \citet*{liebert94} spectroscopy but they now seem to be re-admitted as definite 
cluster member, based on the astrometric analysis of 
\citet{platais11}.\footnote{A further star, B10, is questioned as a field member 
by \citet{platais11}, but see Paper II.}
These alleged cases warn, however, of the possible amount of contamination 
by field interlopers.

\section{Cluster spectral synthesis}

As a final, and possibly most instructive excercise, we could rely on the 
derived set of ($\log L, \log T_{\rm eff}, \log g$) fundamental parameters of
our {\it bona fide} cluster sample to sum up the expected spectrophotometric 
contribution of each star such as to obtain the integrated SED of the cluster 
as a whole. 

\begin{figure}
\centerline{
\psfig{file=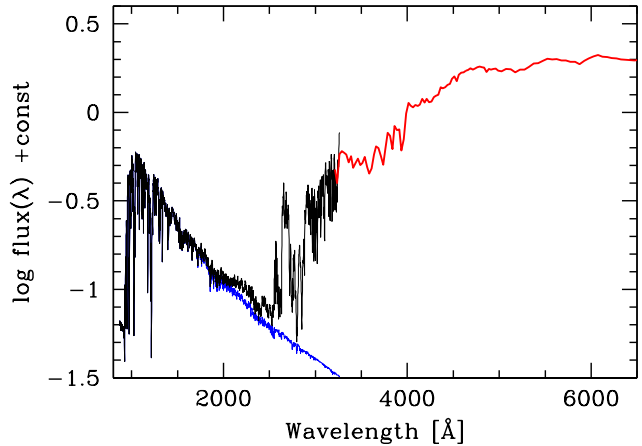,width=\hsize}
} 
\caption{
The synthetic SED of NGC~6791. The integrated spectrum is obtained by summing up the
contribution of our {\it bona fide} star sample of 4739 entries. 
Mid-UV ($\lambda \la 3200$~\AA) spectral synthesis has been carried out at high 
resolution (2~\AA\ FWHM) by attaching each star the corresponding {\sc Uvblue} synthetic 
spectrum with the same fundamental parameters (black line). Same procedure has been adopted 
at longer wavelength (red line), by matching the \citet*{kurucz93} {\sc Atlas9} library
at low resolution ($\sim 25$~\AA). The selective contribution of hot ($T_{\rm eff} \ge 10,000$~K)
is singled out (blue line) showing that these stars are the prevailing contributors to the
striking UV upturn in this SED. According to our estimate (see eq.~\ref{eq:uvtot})
this feature collects about 1.7{\scriptsize $\pm 0.4$}\% of the total bolometric luminosity 
of the cluster.
\label{f5}}
\end{figure}

This task has been first carried out at low spectral resolution especially addressing 
the optical/infrared wavelength range. We made use, in this regard, of the original 
{\sc Atlas9} grid of \citet*{kurucz93} synthetic stellar spectra, sampling 
wavelength at 25-50~\AA\ steps. 
In addition, to gain a better view of the UV distinctive 
properties of the cluster stellar population, we also carried out a high-resolution
(2~\AA\ FWHM) spectral synthesis between 870 and 3300~\AA\ by matching stars 
with the appropriate {\sc Uvblue} spectral grid. As a general
procedure in both cases, each star has been located within the relevant grid 
according to its fundamental parameters, and the corresponding synthetic SED 
has been attached to it providing to normalize 
flux density such as to match the value of the bolometric luminosity.
By summing up all the 4739 entries brighter than $B = 21.5$, we eventually led
to an estimate of the integrated SED of NGC~6791. Our results are displayed
in Fig.~\ref{f5}.
 
The synthetic SED opens to a number of interesting applications.
Besides computing broad-band colors along the entire wavelength range, we could 
also estimate narrow-band spectrophotometric indices, either via direct integration 
on the synthetic spectrum, or by means of the so-called fitting-function procedure
\citep{buzzoni92,worthey94}. 
As for first case, the \citet{fanelli90} mid-UV indices have been obtained,
providing to degrade our spectral resolution to 6~\AA\ (FWHM) to match the standard 
system definition \citep[see also][for details]{chavez07}.
In addition, a relevant subset of Lick narrow-band indices has been obtained by relying
on two independent schemes of index fitting functions, according to 
\citet{buzzoni92,buzzoni94} and \citet{worthey94}. We therefore dealt with
the integrated strength of the H$\beta$ Balmer line, the popular Magnesium 
absorption features (Mg$_2$ and Mg$_b$ indices), and three recognized Fe{\sc i} blends 
around 5000~\AA\  (Fe5015, Fe5270, and Fe5335 indices).
A direct assessment of the 4000~\AA\ Balmer break has been possible, as well, through 
the two indices $\Delta$, originally defined in mag scale by \citet{brodie86} and \citet{brodie90}, 
and $D4000$, in terms of flux ratio as from \citet{bruzual83} (see also \citealp{hamilton85} 
and \citealp{gorgas99}).

For reader's better convenience, the full set of broad-band synthetic colors for NGC~6791,
including the Johnson/Cousins, Gunn, {\sc Galex} photometric systems and Balmer-break indicators
is summarized in Table~\ref{tab:t1}. The narrow-band spectrophotometric indices are
collected, instead, in Table~\ref{tab:t2}.

\begin{table}
\begin{center}
\footnotesize
\caption{Broad-band magnitude and color properties from spectral synthesis of 
NGC~6791$^{(a)}$ compared to standard elliptical galaxies\label{tab:t1}}
\begin{tabular}{lrrl}
\tableline
\\
 & NGC~6791 & Ellipticals$^{(b)}$ &  System ref. \\
\tableline
\\
M$_{\rm bol}$ & --6.29\phantom{1111} &  \nodata\phantom{1111}    & $~~\quad$---\\
Bol--V & --0.69\phantom{1111} &  \nodata\phantom{1111}           & Johnson \\ 
U--B   &  0.60\phantom{1111}  & 0.56\phantom{1111} & $\qquad$" \\ 
B--V   &  0.97\phantom{1111}  & 0.97\phantom{1111} & $\qquad$" \\ 
V--R   &  0.86\phantom{1111}  & 0.86\phantom{1111} & $\qquad$" \\ 
V--I   &  1.50\phantom{1111}  & 1.63\phantom{1111} & $\qquad$" \\ 
V--J   &  2.08\phantom{1111}  & 2.35\phantom{1111} & $\qquad$" \\ 
V--H   &  2.74\phantom{1111}  & 3.05\phantom{1111} & $\qquad$" \\ 
V--K   &  2.90\phantom{1111}  & 3.28\phantom{1111} & $\qquad$" \\ 
V--R$_c$ & 0.60\phantom{1111} & 0.61\phantom{1111} & Johnson/Cousins \\ 
V--I$_c$ & 1.18\phantom{1111} & 1.31\phantom{1111} & $\qquad$" \\ 
M$_g$  & --5.38\phantom{1111} & \nodata\phantom{1111}                   & Gunn \\ 
g--r   &  0.45\phantom{1111}  & 0.40\phantom{1111} & $\qquad$" \\  
g--i   &  0.67\phantom{1111}  & 0.70\phantom{1111} & $\qquad$" \\  
g--z   &  0.79\phantom{1111}  & 0.96\phantom{1111} & $\qquad$" \\ 
1550-V &  2.44{\tiny $\pm 0.25$} & $>2.17${\tiny $\pm 0.15$} & \citet{burstein88}\\
FUV-V  &  5.22{\tiny $\pm 0.25$} & $>4.95${\tiny $\pm 0.15$} & {\sc Galex}$^{(c)}$ \\
NUV-V  &  5.01{\tiny $\pm 0.25$} & $>4.56${\tiny $\pm 0.15$} & $\qquad$" \\
\\
\multicolumn{3}{l}{Balmer break indices:} & \\
$\Delta$ & 0.36\phantom{1111}   &  \nodata\phantom{1111}         & \citet{brodie86}\\     
$D4000$ & 1.82\phantom{1111}     & 2.16{\tiny $\pm 0.24$} & \citet{bruzual83}$^{(d)}$\\   
\tableline
\multicolumn{4}{l}{$^{(a)}$Reddening and distance modulus according to}\\
\multicolumn{4}{l}{$\quad$\citet{twarog09}; error bar for the UV colors from}\\
\multicolumn{4}{l}{$\quad$Poissonian number fluctuation of bright hot stars of Fig.~\ref{f3}.}\\
\multicolumn{4}{l}{$^{(b)}$Johnson colors are from \citet{yoshii88};}\\
\multicolumn{4}{l}{$\quad$Cousins and Gunn colors from \citet{fukugita95};}\\
\multicolumn{4}{l}{$\quad$(1550-V) and {\sc Galex} average colors for the}\\
\multicolumn{4}{l}{$\quad$four strongest UV-upturn galaxies in Fig.~\ref{f7};}\\
\multicolumn{4}{l}{$\quad$Balmer-break index $D4000$ from \citet{hamilton85}.}\\
\multicolumn{4}{l}{$^{(c)}${\sc Galex} colors in AB magnitudes.}\\
\multicolumn{4}{l}{$^{(d)}$Index value as a flux ratio.}\\
\end{tabular}
\end{center}
\end{table}

\begin{table}
\begin{center}
\caption{Mid-UV, Balmer-break and selected Lick spectrophotometric indices of NGC~6791\label{tab:t2}}
\begin{tabular}{lrclrl}
\tableline
\\
\multicolumn{2}{l}{Mid-UV indices:$^{(a)}$} &  & \multicolumn{3}{l}{Lick indices:} \\ 
\multicolumn{2}{l}{\hrulefill} &               & \multicolumn{3}{c}{\hrulefill} \\    
Fe{\sc ii} 2332  & 0.123 &		       & H$\beta$   & 1.92 & \AA \\	      
Fe{\sc ii} 2402  & 0.145 &		       & Fe5015     & 7.44 & \AA \\	      
BL 2538 	 & 0.374 &		       & Mg$_2$     & 0.246 & mag \\	      
Fe {\sc ii} 2609 & 1.328 &		       & Mg$_b$     & 4.28 & \AA \\	     	      
BL 2720 	 & 0.410 &		       & Fe5270     & 3.51 & \AA \\	     	     
BL 2740          & 0.523 &		       & Fe5335     & 3.16 & \AA \\	     	      
Mg {\sc ii} 2800 & 0.668 &		       &            &      &     \\	     	      
Mg {\sc i} 2852  & 0.700 &		       &            &      &     \\ 	     		       
Mg wide 	 & 0.343 &		       &            &      &     \\ 	     	       
Fe {\sc i} 3000  & 0.286 &		       &            &      &     \\ 	     	      
BL 3096          & 0.170 &		       &            &      &     \\ 	     	      
2110/2570	 & 0.132 &		       &            &      &     \\ 
2600-3000	 & 0.969 &		       &            &      &     \\
2609/2660	 & 1.328 &		       &            &      &     \\
2828/2921	 & 0.778 &		       &            &      &     \\
S2850	         & 1.277 &		       &            &      &     \\
S2850L           & 1.447 &		       &            &      &     \\
\tableline
\end{tabular}
\end{center}
$^{(a)}$Index values in mag, as defined by \citet{fanelli90} and \citet{chavez07}\\
\end{table}

A first direct estimate of the integrated color of NGC~6791 has been attempted by \citet{kinman65},
who probed the central region of the cluster obtaining (after correction to our reddening 
scale) $(B-V)_o^{\rm tot} = 0.96$, in perfect agreement with our synthesis
output. An isochrone approximation of the same data, and with the same reddening offset, 
led \citet{xin05} to obtain $(B-V)_o^{\rm tot} = 1.07$, while \citet{lata02}, from a coarser
set of observations, propose $(B-V)_o^{\rm tot} = 1.1 \pm 0.2$ for the cluster. Both these sources, 
however, tend to be biased against bluer objects. Based on the \citet{kaluzny95} and
\citet{stetson03} $BV$ database we carried out different experiments by summing up stars in each
catalog after field-stars cleaning according to the color selection scheme of Sec.~2. 
A more realistic figure, still adopting our $E(B-V)$ color excess, can be envisaged for the 
integrated color in the range $(B-V)_o^{\rm tot} \simeq 0.95 \pm 0.03$ for \citet{kaluzny95}
and $(B-V)_o^{\rm tot} \simeq 1.00\pm 0.02$ for \citet{stetson03}.

According to the H-R transformation, the bolometric emission of the integrated
SED of Fig.~\ref{f5} amounts to $M_o^{\rm bol} = -6.29$, which implies 
a total luminosity of the sampled cluster population of 
\begin{equation}
L_{\rm 6791}^{\rm bol} = 10^{-0.4\,(M_o^{\rm bol}-M_\odot^{\rm bol})} = 25350~L_\odot,
\label{eq:ltot}
\end{equation}
assuming $M_\odot^{\rm bol} = 4.72$ from \citet{portinari04}.
According to our total $V$-band magnitude $M^V_{\rm tot} = -5.60$, cluster luminosity 
results about a factor of two brighter than \citet*{kinman65} original output (once converting 
to our reddening scale and distance modulus), but one has to consider, in this regard, that we 
are sampling a factor of four wider field. Again, with our prescriptions for
distance and reddening correction, the previous experiments with the \citet{kaluzny95} 
and \citet{stetson03} $BV$ stellar sets provided us with a value of $M^V_{\rm tot} = -5.95 \pm 0.07$,
and $-5.62 \pm 0.03$, respectively, in far better agreement with our prediction, once 
considering the slightly different sampled area.

\begin{figure}
\centerline{
\psfig{file=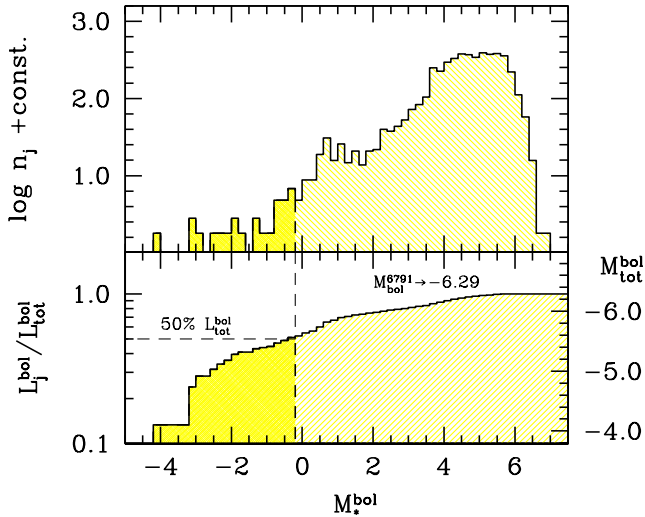,width=\hsize}
} 
\caption{
{\it Upper panel} - The bolometric luminosity function of the NGC~6791 stellar population
according to the c-m diagram of Fig.~\ref{f3}.
{\it Lower panel} - The cluster integrated magnitude obtained by summing up stars with 
increasingly fainter bolometric luminosity. An asymptotic value of $M^{\rm bol}_{tot} = -6.29$
is reached, when including all the 4739 stars in our sample. Note the outstanding contribution
of the few bright stars with negative value of $M^{bol}_*$ (some 30 objects in total), 
which provide about 50\% of cluster total luminosity.
\label{f6}}
\end{figure}

Although a prevailing fraction of the cluster total luminosity certainly comes from the few
bright giant stars in our sample, one may argue, however, that our estimate of $L_{\rm 6791}^{\rm bol}$ 
is a somewhat lower limit as {\it i)} we are missing the contribution of the faintest 
MS stars and {\it ii)} we might be sampling just a fraction of the total 
cluster, according to our observed field.
A preliminary estimate of point {\it (i)} can be attempted by relying on the
combined analysis of the luminosity function and the cumulative luminosity contribution
of the star counts vs.\ absolute stellar luminosity. This is shown in Fig.~\ref{f6}.
One may guess from the figure that sampling completeness of our 
luminosity function safely extends down to $M_*^{\rm bol} \la +5.5$.
Even assuming star counts to steadily increase with the same trend also at fainter 
magnitudes\footnote{An intrinsic flattening of the luminosity function just fainter than the Turn Off 
point (i.e. $M_*^{\rm bol} \simeq +4.0$) is however confirmed also by the results of \citet{kaluzny95}.} 
we verified that the inferred $L_{\rm 6791}^{\rm bol}$ value would increase by 15\% at most.

Concerning the spatial sampling, as for point ({\it ii}) of our {\it caveat},
we made use of the $V$ data from \citet{stetson03}, collected over a larger
field ($\sim 20^\prime \times 20^\prime$) compared to our observations, to derive an 
estimate of the lost luminosity fraction. From these data, after fitting the star 
distribution with a \citet{king66} profile we obtain a photometric (projected) and tidal
radius for the cluster, respectively of $R_c = 4.75 \pm 0.29$, and 
$R_t = 20.57 \pm 7.48$~arcmin,\footnote{Note that the large error of our $R_t$ estimate
are due to a mild spatial extrapolation of the \citet{stetson03} data.}
in nice agreement with the recent study of \citet{platais11}. 
With these shape parameters, according to the $V$ luminosity profile,
we estimate that about 95\% of cluster luminosity has been effectively
collected by our observations.

\section{NGC~6791 as a proxy of UV-upturn elliptical galaxies?}

Based on the resolved H-R diagram of Fig.~\ref{f3}, the contribution of the 
different stellar evolutionary stages to the integrated SED of the cluster can easily be 
disaggregated. As an outstanding feature, in this regard, one sees from
Fig.~\ref{f5} that the hot stellar component with $T_{\rm eff} \ga 10,000$~K 
is by far the prevailing contributor to the UV luminosity of the cluster
shortward of 2500~\AA. Actually, these stars are the main responsible of the striking 
``UV upturn'' in the spectrum of NGC~6791. A so close resemblance with the corresponding
feature that sometimes marks the SED of elliptical galaxies evidently calls for
special analysis of our data, in order to assess how confidently we can retain
NGC~6791 as a suitable ``proxy'' of UV-enhanced ellipticals.

The strength of the UV upturn for the cluster is probed in Table~\ref{tab:t1}
by the classical $(1550-V)$ color, as originally defined by \citet{burstein88}.
A nearly equivalent definition can be obtained in an updated AB mag scale based
on the {\sc Galex} photometric system, as $(FUV-V) \simeq (1550-V) +2.78$
\citep[see, e.g.][]{neff08}, as reported in the same table.
According to our synthesis output, hot stars supply a total of 
430~$L_\odot$ so that the relative contribution of the UV emission 
to the total bolometric luminosity of the cluster is 
\begin{equation}
{{L^{\rm UV}_{\rm hot *}} \over {L^{\rm bol}_{6791}}} = {430\over 25350} = 
\begin{array}{l}
 \\
0.017,\\
\,\quad^{\pm 4}
\end{array}
\label{eq:uvtot}
\end{equation}
where error bar has been estimated from the Poissonian number fluctuation of the 12 
bright hot stars of Fig.~\ref{f3}.

\begin{figure}
\centerline{
\psfig{file=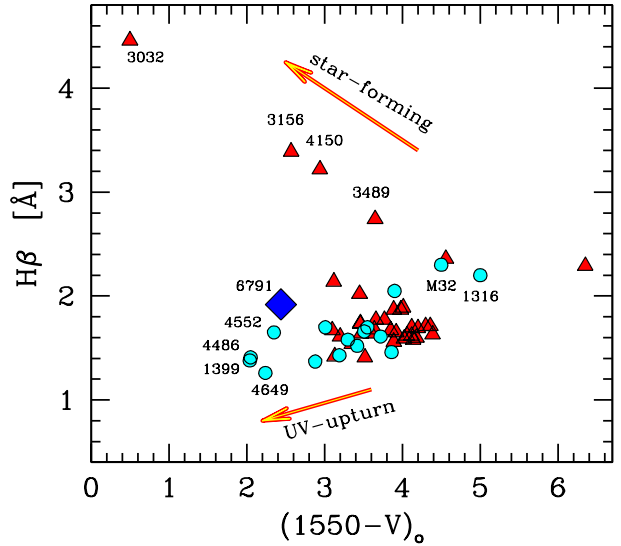,width=\hsize}
}
\caption{A comparison of the NGC~6791 location in the $H\beta$ index vs. $(1550-V)$ UV color
(big romb marker, as labelled) is carried out with a set of elliptical galaxies along a 
range of evolutionary cases. The \citet{bureau11} {\sc Sauron} database of 41 ellipticals 
with available {\sc Galex} UV photometry is displayed (triangles) together with the
original {\sc Iue} sample of \citet{burstein88}, as recompiled by \citet{buzzoni08} 
(18 galaxies with available Lick indices) (solid dots).
Some reference objects are singled out and labelled for better clarity (see text for
a discussion). A clear ``$>$''-shaped pattern of galaxy distribution is in place, with
mildly star-forming systems displaying a stronger ($\ga 2$\AA) $H\beta$ index, contrary to 
UV-upturn galaxies, which supply the same UV emission with a much shallower $H\beta$
absorption. Note the resemblance of NGC~6791 with the case of galaxy NGC~4552.
\label{f7}}
\end{figure}

To properly set this figure in the more general extragalactic framework, in 
Fig.~\ref{f7} we compare the NGC~6791 output with a set of elliptical galaxies spanning 
a wide range of evolutionary cases. In particular, the original {\sc Iue} sample 
of \citet{burstein88}, recently recompiled by \citet{buzzoni08} (18 galaxies with 
available Lick indices), has been extended by including the {\sc Sauron} database of 
ellipticals with available {\sc Galex} UV photometry, as discussed by \citet{bureau11}
\citep[41 galaxies, whose 5 objects in common with][]{buzzoni08}.

A more effective evolutionary characterization of the whole galaxy sample can be
done by contrasting the $(1550-V)$ color with the H$\beta$ equivalent width, as probed 
by the corresponding Lick index.
Being especially sensitive to A-F stars \citep{gorgas93,buzzoni95,buzzoni09}, $H\beta$ is 
an effective tracer of the Turn-Off temperature in the c-m diagram of intermediate-age 
(a few Gyr, or so) stellar populations. A stronger $H\beta$ absorption,
combined with enhanced UV luminosity is therefore an unequivocal sign of moderate
but recent star-formation activity in a galaxy, while a shallower feature always 
points to the presence of an underlying old (quiescent) stellar population.

The ``$>$''-shaped trend of the galaxies in Fig.~\ref{f7} actually summarizes the
general picture, with a clear sequence of star-forming ellipticals 
\citep[i.e. NGC~3032, 3156, 4150 etc., see e.g.][for a discussion]{temi09},
which diagonally crosses the plot, with decreasing $H\beta$ and $(1550-V)$, and 
ending up in the lower right corner with the relevant case of NGC~1316 
\citep[a fresh post-merger object, see e.g.][]{goudfrooij01} and M32.
With further decreasing the Balmer index (roughly for $H\beta \la 2$~\AA), one sees 
that the UV upturn begins to take shape especially for the group of giant
ellipticals, namely NGC~4552, 4486, 1399, and 4649, characterized by an increasingly 
``blue'' $(1550-V)$ color. 
In this framework, no doubt the case of NGC~6791 fits well with the quiescent ellipticals
with the strongest UV upturn, although one has to notice the slightly larger $H\beta$,
in consequence of the relatively younger age of the cluster compared to the galaxy 
stellar populations.

Assuming the UV slope not to vary so much among ellipticals,
one can rely on the case of NGC~6791 to set a scale relationship between
the $(1550-V)$ color, and the relative strength
of galaxy UV emission, compared to the global bolometric luminosity.
According to the color definition, and relying on the figures of Table~\ref{tab:t1} we 
simply obtain\footnote{A similar relation can also be derived 
for the {\sc Galex} $(FUV-V)$ color such as 
$(L^{\rm UV}/ L^{\rm bol}_{gal}) = 2.06\,10^{-0.4\,(FUV-V)}$.}
\begin{equation}
{{L^{\rm UV}} \over {L^{\rm bol}_{gal}}} = 0.16\,10^{-0.4\,(1550-V)}.
\label{eq:uvpercent}
\end{equation}
According to the galaxy distribution of Fig.~\ref{f7}, UV-upturn ellipticals 
are expected to emit between 1 and 2.5\% of their bolometric luminosity in the 
ultraviolet.

\begin{figure}
\centerline{
\psfig{file=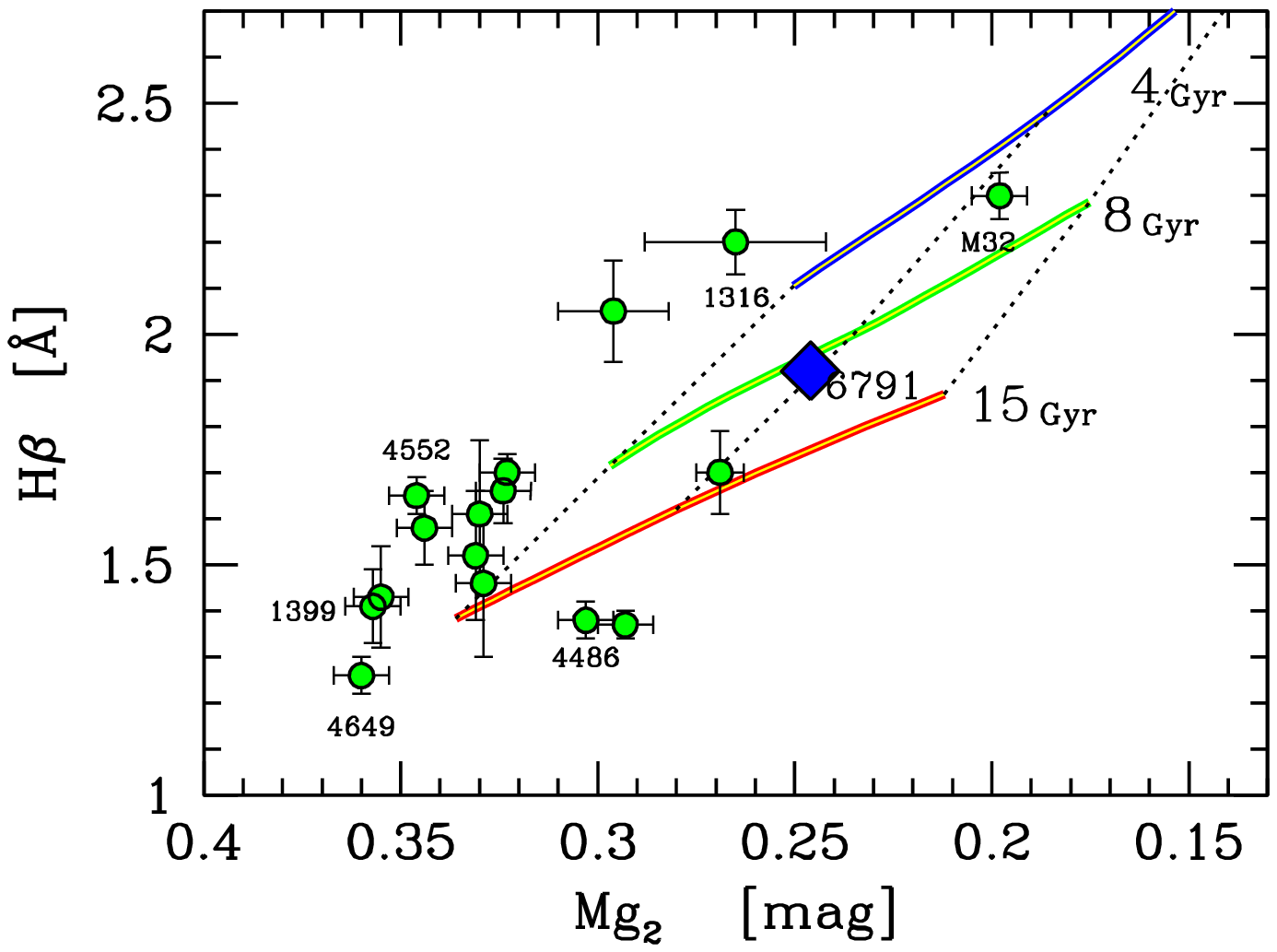,width=\hsize}
}
\caption{
The \citet{buzzoni08} compiled sample of elliptical galaxies is displayed 
in the classical diagnostic plane of the $H\beta$ vs.\ Mg$_2$ Lick indices, together
with the corresponding location for NGC~6791 (big romb marker). Observations are compared 
with the \citet{buzzoni89}
and \citet{buzzoni95} population synthesis models (with $Y = 0.25$) for an age of 
4, 8 and 15 Gyr as labelled (solid lines), and for a metallicity parameter 
$[Fe/H] = -0.5, 0.0$, and +0.5, in the sense of increasing Mg$_2$ (dotted line envelope).
In spite of its recognized value of $[Fe/H] \sim +0.4$, NGC~6791 appears to match here 
only a marginally super-solar metallicity due to its relatively low value of 
Mg$_2$. This is in consequence of the scanty population of RGB+AGB stars,
a feature that we may adscribe to the speeded-up RGB evolution induced by Helium 
overabundance. See text for a discussion.
\label{f8}}
\end{figure}

A more general view of the stellar bulk of NGC~6791 can be gained 
in the domain of classical broad-band colors. A comparison with standard ellipticals is
proposed in Table~\ref{tab:t1}, after compiling Johnson, and Cousins/Gunn mean colors from 
the original works of \citet{yoshii88} and \citet{fukugita95}, respectively.
We also took advantage of \citet*{hamilton85} analysis of the $D4000$ Balmer-break index in
the local and cosmological framework to derive a mean representative value of the index for
present-day ellipticals (see Table~2 therein). This is reported in Table~\ref{tab:t1}, as well,
together with its observed standard deviation.

The UV-upturn feature is quantified in the table from the data of Fig.~\ref{f7} by averaging
the $(1550-V)_o$ color of the four most UV-enhanced ellipticals in the plot, namely
NGC~1399, 4486, 4552 and 4649. For these galaxies we also assumed a {\sc Galex}
color $(FUV-NUV) = 0.39$ from the reference SED of \citet{buzzoni08}. According to \citet{yi11},
only some 5\% of present-day early-type galaxies do actually display the UV-upturn phenomenon.
For this reason, our figures in Table~\ref{tab:t1} should be taken as lower limits to the
allowed color range for standard ellipticals in the local Universe.

As far as optical colors are concerned, we see that NGC~6791 strictly matches the photometric properties
of standard ellipticals. A systematic difference begins to appear, on the contrary, when
moving to longer wavelength in the near infrared, with increasingly brighter $JHK$ magnitudes
relative to the $V$ luminosity (``redder'' colors) for ellipticals with respect to the open
cluster. 
An interesting difference can also be noticed for the 4000~\AA\ Balmer jump, which is significantly 
shallower for the cluster with respect to the average elliptical. When specifically contrasted with 
UV-upturn galaxies, like NGC~4552 ($D4000 = 2.30$) and NGC~4486 ($D4000 = 2.06$), we see from  
\citet{hamilton85} that in any case galaxy $D4000$ index exceeds the NGC~6791 estimate. 
As the $D4000$ feature mainly collects the photometric properties of the Turn Off stars, 
increasing in value along the early- late-type spectral sequence 
\citep[see the instructive Fig.~6 of][]{hamilton85}, we are
inclined to interpret this difference as a sign of the younger age of NGC~6791.

\begin{table*}
\begin{center}
\caption{Star number counts across the H-R diagram of NGC~6791 and implied stellar lifetime and fuel
consumption\label{tab:t3}}
\begin{tabular}{lrrrrcc}
\tableline
\\
Stage  &  number   & implied lifetime  & \multicolumn{2}{c}{theoretical lifetime}   &   $L_j/L_{\rm tot}$ & H-equivalent \\
       &  of stars &      & {\sc Basti} & {\sc Padova} &                &  fuel [$M_\odot$] \\ 
\tableline
\\
MS              & $\gg 3307$ & \nodata                    & \nodata  & \nodata  & 0.10 &  \nodata \\
SGB             & 881	     & 1762{\tiny $\pm 59$} Myr  & 2390 Myr & 1660 Myr  & 0.08 & 0.039{\tiny $\pm 0.001$} \\
RGB+AGB         & 333	     &  666{\tiny $\pm 37$} Myr   & 735 Myr  & 633 Myr  & 0.64 & 0.304{\tiny $\pm 0.017$} \\
RHB+EHB         & 45+12	     &  114{\tiny $\pm 15$} Myr   & 107 Myr  & 113 Myr  & 0.10 & 0.047{\tiny $\pm 0.006$} \\
WD              & $\gg 10$   & \nodata  	          & \nodata  & \nodata  & 0.00 & \nodata \\
Unclass.        & 151	     & \nodata                    & \nodata  & \nodata  & 0.08 & 0.040{\tiny $\pm 0.003$} \\
                & \hrulefill &  	&	&    & \hrulefill &  \hrulefill        \\
Total           &  4739      &  	&	   &            &    1.00  &   0.43~M$_\odot${\tiny $\pm 0.01$} \\
\tableline
\\
BRGB{\tiny +AGB}& 35         &   70{\tiny $\pm 12$} Myr   &  67 Myr  &  59 Myr  &      &       \\
R$^\prime$ =  & 57/35  &  1.63{\tiny $\pm 0.50$}\phantom{Myr} & 1.60 {\tiny (Y=0.30)} & 1.92 {\tiny (Y=0.34)} &      &        \\
\tableline
\end{tabular}
\end{center}
\end{table*}

Finally, a comparison of the two most popular Lick indices, namely H$\beta$ and Mg$_2$,
is attempted in Fig.~\ref{f8}. As well known \citep{buzzoni95,buzzoni95b} 
this combination allows a simple and quite effective ``taxonomy'' of the H-R diagram 
of a stellar population as, while H$\beta$ strength probes the Turn-Off location of 
the population (and therefrom its age), the integrated Mg$_2$ index better responds 
to the red-giant branch (AGB+RGB) contribution.
When compared to its location in Fig.~\ref{f7}, NGC~6791 moves now in the 
Lick plane toward the region of low-mass metal-poorer ellipticals, like M32. 
When matched with standard population synthesis models \citep[e.g.][]{buzzoni89,buzzoni95} 
the H$\beta$ strength consistently confirms for the cluster an age about 8~Gyr, 
while Mg$_2$ points to just a marginally supersolar metallicity. According to our
discussion in Sec.~3.1, this behavior may be the natural effect of a larger 
$R$ parameter, as induced by an enhanced Helium abundance. A coarser number of
bright RGB+AGB stars would reduce in fact the integrated Mg$_2$ index and could also
give reasons of the sensibly ``bluer'' infrared colors, thus mimicking the effect of a lower 
metallicity.

\section{Implied stellar lifetimes}

\begin{figure}
\centerline{
\psfig{file=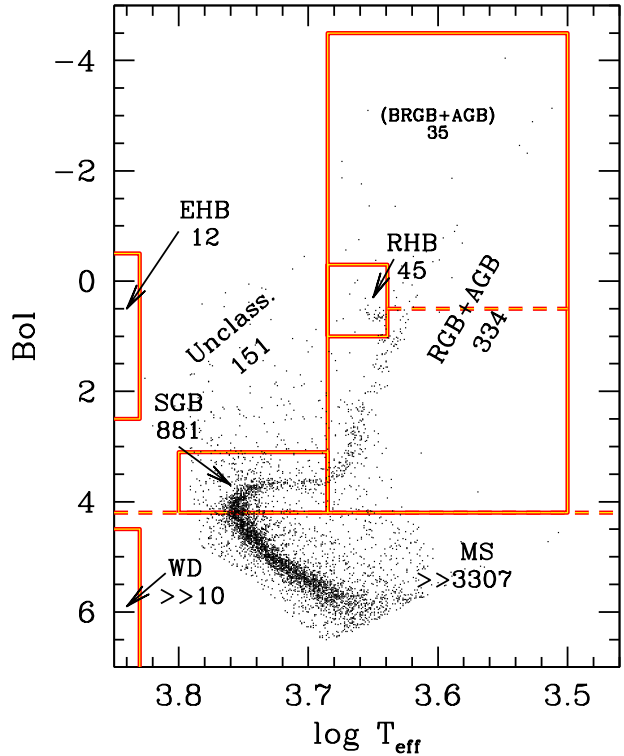,width=0.96\hsize}
}
\caption{
A zoomed-in sketch of the cluster H-R diagram with the selection scheme to pick up
stars in the different evolutionary stages. The relevant star number counts (proportional
to the evolutionary lifetime for PMS stars, according to eq.~\ref{eq:life} are reported 
in the different boxes, as summarized in better detail also in Table~\ref{tab:t3}.
\label{f9}}
\end{figure}

Star number counts across our resolved H-R diagram of 
NGC 6791 can also be used to infer the absolute lifetime of the different
post-Main Sequence (PMS) evolutionary stages. This can be done by relying on the
the so-called ``Fuel consumption theorem'' (FCT) \citep{rb86} which, for every ``j-th''
PMS phase, allows us to probe its implied lifetime ($\tau_j$) just in terms of 
the number ($n_j$) of observed stars that currently sample it in the cluster diagram. 
Following \citet{rb86}, once accounting for the total collected luminosity of NGC 6791, 
we can write
\begin{equation}
n(j) = {\cal B}\, L^{\rm bol}_{6791}\, \tau(j)
\label{eq:b}
\end{equation}
In the equation, the parameter ${\cal B}$ is the so-called ``specific evolutionary flux'' 
and, according to stellar population synthesis models \citep[see, e.g.][]{buzzoni89,maraston98}, it 
can safely be set to ${\cal B} \sim 2\,10^{-11}$~L$_\odot^{-1}$yr$^{-1}$. 
By recalling eq.~(\ref{eq:ltot}), we eventually obtain
\begin{equation}
\tau(j) \simeq 2.0~10^6\,n(j) \qquad{\rm [yr]}.
\label{eq:life}
\end{equation}
In our sample one has therefore to expect, on average, one star per PMS 
evolutionary step of 2~Myr.
Star number counts have been carried out for the NGC~6791 stellar population
according to the scheme of Fig.~\ref{f9}. Our results are summarized in 
col.\ 2 of Table~\ref{tab:t3} together with the implied lifetime (col.\ 3), as from
eq.~(\ref{eq:life}).

Although within the large uncertainties due to the low statistics, a close inspection 
of our data and a match with the several c-m diagrams available in the literature 
\citep[see, in particular][on this subject]{liebert94} 
allows us a preliminary estimate of the \citet{iben68} $R$ parameter
to probe the Helium content of the NGC~6791 stellar population.
More properly, as a variant in case one cannot firmly discriminate between
AGB and bright RGB stars, also an alternative parameter $R^\prime = N_{\rm HB}/N_{\rm GB}$ 
can be tried, where $N_{\rm HB} = N_{\rm EHB}+N_{\rm RHB}$ and 
$N_{\rm GB} = N_{\rm BRGB}+N_{\rm AGB}$ includes both the RGB tail brighter than the RHB 
luminosity level, and the AGB contribution. According to Table~\ref{tab:t3}, we have 
$R^\prime = 57/35 = 1.63${\scriptsize $\pm 0.50$}, where the error bar comes from a logarithmic 
differentiation, so that $dR'/R' \sim dN_{\rm HB}/N_{\rm HB} + dN_{\rm GB}/N_{\rm GB}$.

By relying on the \citet{buzzoni83} calibration, one may eventually figure out a Helium abundance
for the cluster of $Y_{\rm 6791} = 0.30${\scriptsize $\pm 0.04$},
in close agreement with the results of \citet{brogaard11} based on high-res
spectroscopy of detached eclipsing binaries.
This implies an enrichment ratio $\Delta Y/\Delta Z \sim 2$.

Although moving from opposite physical arguments, our picture may actually complement the 
original interpretative scheme proposed by \citet{kalirai07}, as we have been 
anticipating in Sec.~3.1.
Both the effect of enhanced mass loss and Helium abundance act, in fact, in the sense of 
increasing the $M_c/M_e$ ratio of PMS stars either by decreasing the envelope 
size (mass loss) or by increasing the He core mass (He overabundance).
An enhanced $M_c/M_e$ ratio favors post-RGB stars to anticipate their pathway to the 
white-dwarf final stage, in some cases by directly relocating in the high-temperature region 
of the H-R diagram both in the form of EHB or ``hot flasher'' 
stars \citep{dcruz96,castellani06b,miller08,yaron08}.
Besides their synergic behavior to produce hotter HB stars, one major 
difference between the two evolutionary scenarios is that, contrary to mass loss, a change in 
Helium abundance directly acts on the evolutionary lifetime as it affects the nuclear clock of stars
along the whole He-burning phases.

A useful comparison of our output can be done with predictions from stellar evolution 
theory. Again, for this match we relied on the two reference sets of stellar tracks from 
the {\sc Basti} \citep{pietrinferni04} and {\sc Padova} \citep{bertelli08} databases 
assuming a very similar (solar-scaled) chemical composition, namely $(Z, Y) = (0.04, 0.30)$
for the {\sc Basti} models and $(Z, Y) = (0.04, 0.34)$ for the corresponding 
{\sc Padova} ones. Considering cluster age and chemical composition, a Turn-Off
stellar mass of $M_{\rm TO} \simeq 1.1\, M_\odot$ has to be expected. This
value is also consistent with the direct estimate of the primary stellar component
of the detached binary system V20, almost exactly placed at the Turn Off point,
and for which \citet{grundahl08} indicate a mass of $1.074~M_\odot$.
The resulting theoretical lifetime for PMS evolution is displayed in cols.\ 4 
and 5 of Table~\ref{tab:t3}.

In addition, we also added to the table (see col.\ 6) the relative bolometric contribution 
to cluster luminosity as from the observed stars in the different evolutionary stages. 
This ratio is actually the essence of the FCT, and gives a 
direct hint of the nuclear fuel consumed by stars along their PMS evolution.
By simply rearranging eq.~(\ref{eq:b}) we can write in fact
\begin{equation}
(n\,\overline{{\cal \ell}})_j = {\cal B}\, L_{\rm tot}\, (\tau\,\overline{{\cal \ell}})_j,
\label{eq:nt}
\end{equation}
where $\overline{{\cal \ell}}(j)$ is the mean reference luminosity of stars
along their j-th evolutionary phase. Allover, this phase will provide a total emission  
$L_j = (n\,\overline{{\cal \ell}})_j$ and a relative contribution to the global
cluster luminosity of
\begin{equation}
{L_j\over L_{\rm tot}} = {\cal B}\, {\cal F}_j.
\label{eq:fct}
\end{equation}
In the equation, ${\cal F}_j =  (\tau\,\overline{{\cal \ell}})_j$ is the fuel consumed 
by a star along its j-th PMS evolutionary stage. After appropriate conversion
\citep[see][for the detailed procedure]{buzzoni11} this figure can also be expressed in 
H-equivalent $M_\odot$, as reported in col.\ 7 of Table~\ref{tab:t3} 
(see also \citealp{renzini88} and \citealp{maraston98} for an instructive application
of this approach to other Galactic globular clusters).

A few interesting considerations stem from the combined analysis of the table entries.
The first, and possible most important one, deals with the implied HB lifetime.
As far as the number counts is concerned, one sees
from the table that a full score, including both the RHB and the EHB stars, is required
to consistently compare with the theoretical expectations from both the {\sc Basti} 
and {\sc Padova} stellar tracks. This confirms therefore that both RHB and EHB stars
are part of the {\it same} physical stage along the evolutionary path.
According to our data, the HB phase for NGC~6791 stars lasts a total of
$(45+12)\times 2 = 114${\scriptsize $\pm 15$}~Myr in quite good agreement with theory.
Along this phase, stars burn a total of 0.047{\scriptsize $\pm 0.006$}~$M_\odot^H$, an output
that has to be compared with the theoretical fuel figures of $0.041~M_\odot^H$ 
and $0.057~M_\odot^H$ for the {\sc Basti} and {\sc Padova} stellar tracks, respectively.
Concerning the claimed presence of a few ($\sim 10$) supplementary HB stars at intermediate 
temperature filling the gap between the RHB clump and the EHB group, as suggested 
by \citet{platais11} but never confirmed before, this result seems unlikely as the implied 
HB lifetime could be hardly explained in terms of standard theory of stellar evolution.

While the energetic arguments about fuel consumption successfully account for star counts 
and the implied stellar lifetime, they could hardly give reason of the peculiar HB morphology 
as observed in NGC~6791, a striking sign of mass bimodality among HB stars. Other intervening 
mechanisms should likely be invoked in this sense to modulate HB morphology, perhaps 
through an ``outside-in'' action on the stellar envelope (thus leaving untouched the stellar clock), 
for instance as a consequence of binary mass transfer \citep{yong00,yi04}. 

As for the bright RGB (BRGB) evolution, models agree to predict 
that stars complete their run from the HB luminosity level up to the RGB tip
in about 51~Myr ({\sc Basti}) or 42~Myr ({\sc Padova}). If one adds further 16-17~Myr 
for stars to accomplish their AGB evolution then, according to eq.~(\ref{eq:life}),
a total of 30-35 bright stars are to be expected in our sample, in fairly good
agreement with what we observe. According to Table~\ref{tab:t3}, the total ``fuel'' 
consumed by stars along their full RGB+AGB path amounts to 0.30{\scriptsize $\pm 0.02$}~$M_\odot^H$,
which is close to the corresponding theoretical figure of $0.25~M_\odot^H$, as confirmed 
both by the {\sc Basti} and {\sc Padova} models.
In force of our previous arguments about the $R^\prime$ ratio, this evidence
points again to a relatively ``quiescent'' evolutionary scenario for
NGC~6791, where the apparent deficiency  in bright red giants seems a natural consequence 
of the enhanced Helium abundance rather than of any harassing effect of mass loss via 
enhanced stellar wind. This argument is further supported by the results 
of \citet{vanloon08}, based on Spitzer observations, who find a definite lack of 
evidence for any circumstellar dust production that would accompany, in case,
enhanced mass loss.

On the other hand, the \cite*{kalirai07} unescapable evidence
for 0.43~M$_\odot$ WDs in the cluster evidently suggests that {\it another} mechanism
must be at work providing an efficient and alternative way for stars to loose
mass. As a further piece of evidence, on this line, one may also recall the recent
works of \citet{bedin08} and \citet{twarog11}, where a similar figure (namely
$\sim 30\% \pm 10\%$) is independently found for the fraction of binary stars in 
NGC~6791. Further evidence for unresolved (close) binary systems among RHB stars
has been also provided by \citet{stello11}, based on asteroseismology observations 
from the Kepler space mission. Such a sizeable presence of binary systems has actually 
been meant by \citet{bedin08} to originate the WD peculiar distribution.

According to Fig.~\ref{f9}, about 3\% of our total {\it bona fide} sample (151 stars)
cannot easily be arranged in our classification scheme. They provide 8\% of our estimated 
cluster luminosity. If simply neglected, as in the most drastic interpretation
supposing to be all spurious interlopers, then we should decrease $L^{\rm bol}_{6791}$ 
accordingly, and this would lead to a correspondingly higher (+8\%) implied PMS lifetime, 
through eq.~(\ref{eq:b}).
So, our estimates in Table~\ref{tab:t3} should be taken, in case, as lower limits.

A thorough assessment of membership probability in this region of the H-R 
diagram \citep[see][]{platais11} may actually lead to envisage a certain fraction of 
field stars. However, one could argue that the evident overdensity of faint stars
(some 30 objects, just above the Turn-Off region about $M_{\rm bol} \sim +3$) is 
certainly to be ascribed to the sizeable population of cluster Blue Stragglers as extensively 
recognized in other literature studies \citep{kaluzny92,rucinski96,landsman98, kalirai07}.
\citet{ahumada07}, in their revised catalog, report 75 blue stragglers candidates for this cluster.

If fully considered in the cluster budget, according to the FCT the 151 ``unclassified'' 
stars would imply a supplementary fuel consumption of roughly $0.04~M_\odot^H$. Therefore, a total
of $0.43~M_\odot$ of Hydrogen (mainly converted to Helium and some metals) has been 
consumed by NGC~6791 stars along their PMS evolution. Interestingly enough, this figure 
is very close to the \citet{kalirai07} estimate of WD mass, so that one may conclude 
that cluster stars reach the end of their PMS path after having fully exhausted (or lost)
their external envelope.

A final consideration deals with the MS contribution, which in our sample amounts to
a scanty 10\% of the total cluster luminosity. This has evidently to be regarded as 
a lower limit, being the faint red dwarf stars off the limit of our detection.\footnote{Note, 
by the way, that part of the faintest undetected population of MS stars in the cluster may
in fact already be accounted for in our photometry via blending effects with brighter objects.
This would actually make ``detected'' stars slightly brighter and redder. A hint in this sense 
is for instance the drift toward ``cooler'' temperatures in the faint-end MS locus of Fig.~\ref{f3} 
when compared with theoretical isochrones.}
In force of the discussion in Sec.~4 (see, in particular, Fig.~\ref{f6} therein)
one could set a quite safe upper limit to the MS luminosity contribution such as 
$L_{\rm MS}/L_{\rm tot} < 0.20$. With this figure, however, even a quick check with the 
theoretical predictions from stellar population synthesis \citep{buzzoni89,buzzoni95} 
firmly points to a dwarf-depleted IMF (that is with a slope consistent or flatter than 
the Salpeter case) for cluster stars.

\section{Summary \& conclusions}
 
In this work we have carried out an analysis of the distinctive evolutionary properties
of the stellar population in the old metal-rich Galactic open cluster NGC~6791.
Our discussion moves from a set of orginal $UB$ observations, which sampled the cluster
field down to $B \sim 22$~mag. In lack of any systematic membership parameters for our 
stars, a statistical cleaning of the field interlopers across the cluster region has 
been adopted through an appropriate magnitude/color selection in the observed 
$B$ vs.\ $(U-B)$ c-m diagram as sketched in Fig.~\ref{f1}. 
To better probe also the few bright red giants in the cluster we complemented our 
$UB$ sample with the 2{\sc Mass} $JK$ output on the same field. The merged database 
of {\it bona fide} cluster stars brighter than $B = 21.5$ eventually consisted of 4739
objects.

Based on the {\sc Uvblue} theoretical library of synthetic stellar spectra \citet{rodriguez05}
we then derived a grid of calibrating relations for $[Fe/H] = +0.4$ such as to link the $(U-B)$ 
color with the effective temperature and the $B$-band bolometric correction (see Fig.~\ref{f2}). 
This allowed us to estimate the fundamental parameters (namely $\log L$, $\log T_{\rm eff}$
and $\log g$) for 4706 available stars with accurate $(U-B)$ color, assuming a color
excess $E(B-V) = 0.125$ and a distance modulus $(m-M)_o =13.07$ after \citet{twarog09}. 
A similar approach has also been carried out for the 94 2{\sc Mass} stars relying on their 
$(J-K)$ color, throught the \citet{buzzoni10} empirical calibration, as in the set of 
eq.~\ref{eq:lt}. The 61 stars of the 2{\sc Mass} sample in common with the $UB$ 
database allowed us to assess the internal accuracy of our calibration procedure.
Accordingly, we concluded that effective temperature and bolometric luminosity
have been retrieved, respectively, within a $\Delta \log T_{\rm eff} \simeq 0.03$~dex and 
$\Delta Bol \simeq 0.2$~mag uncertainty. As a final output of our procedure, 
the H-R diagram of the cluster was obtained. This is shown in Fig.~\ref{f3} and has been the main 
reference for our investigation. The upper MS of the diagram is well matched by the 
{\sc Padova} isochrones of \citet{bertelli08} for $t \simeq 7\pm 1$~Gyr, assuming a 
$(Z,Y) = (0.04, 0.34)$ chemical mix. The Turn Off point appears to be placed at 
$M^{\rm bol}_o = 4.2${\scriptsize $\pm 0.2$}, with an effective temperature 
$T_{\rm eff} \simeq 5700${\scriptsize $\pm 50$}~K, consistent with a spectral type G5V \citep{johnson66}. 

According to the fiducial $\log L$, $\log T_{\rm eff}$ and $\log g$ parameters, we attached each 
star in our sample the appropriate synthetic spectrum eventually obtaining the integrated SED of 
the cluster. In particular, high-resolution (2~\AA\ FWHM) spectral synthesis has been carried out 
for the ultraviolet wavelength region, relying on the {\sc Uvblue} models, while its extension
to longer wavelength was accomplished at lower resolution by means of the \citet{buzzoni89} 
synthesis code, which makes use of \citet*{kurucz93} {\sc Atlas9} grid of model atmospheres 
(see Fig.~\ref{f5}).
As a main output of our synthesis procedure, integrated cluster magnitudes and broad-band colors 
both in the Johnson/Cousins and Gunn systems have been obtained, as summarized in Table~\ref{tab:t1}.
Furthermore, the full set of \citet{fanelli90} mid-UV narrow-band indices were computed from the
high-res SED. These have been complemented with a supplementary set of indices to assess
the 4000~\AA\ Balmer break \citep{brodie86,bruzual83} and the strength of the main spectral 
features in the optical range comprised in the Lick narrow-band index system \citep{worthey94}
(see Table~\ref{tab:t2}).
The sum of all stars in our sample yields a total bolometric magnitude of $M^{\rm bol}_{6791} = -6.29$,
or $L^{\rm bol}_{6791} = 25350~L_\odot$. A residual 15\% luminosity fraction, at most, may have 
been lost in our census residing into the faintest undetected low-MS stars (see Fig.~\ref{f6}).
The corresponding figure for the $V$ band is $M^V_{6791} = -5.60$, or 14300 $V$ solar 
luminosities, assuming from \citet{portinari04} $M^V_\odot = +4.79$.

As far as integrated cluster colors are concerned, according to Table~\ref{tab:t1},
we obtain $(B-V)_{6791} = 0.97$ in farly good agreement with \citet*{kinman65} original
estimate, and with a direct empirical check by summing up stars in the \citet{kaluzny95} and
\citet{stetson03} photometric catalogs, once picking up cluster locus according to our
selection criteria. When compared with the color properties of standard ellipticals
\citep[][see Table~\ref{tab:t1}]{yoshii88,fukugita95}, NGC~6791 appears to be a fairly good 
proxy. To a closer analysis, however, one has to notice significantly 
bluer infrared colors for the cluster, and a lower $Mg_2$ index. All these features are 
clearly reminiscent of a weaker photometric contribution of red-giant stars in our system. 
The younger age of NGC~6791, compared to standard elliptical galaxies, also 
reflects in a slightly stronger $H\beta$ absorption and a shallower 4000~\AA\ Balmer 
break, as reported in Table~\ref{tab:t2} (see also Fig.~\ref{f8}).

The UV properties of the integrated SED have been assessed both in terms
of the classical $(1550-V)$ color, as originally defined by \citet{burstein88}, and
the updated $(FUV-V)$ and $(NUV-V)$ AB colors to match the {\sc Galex} photometric system.
Once compared with extragalactic observations of elliptical galaxies, NGC~6791 appears to
share the UV properties of the most active UV-upturn ellipticals, like NGC~4552 and NGC~4486
(see Fig.~\ref{f7}) with a fraction of 1.7{\scriptsize $\pm 0.4$}\% of its bolometric 
luminosity emitted in the ultraviolet wavelenght range, shortward of 2500~\AA\ (see Fig.~\ref{f5}).
This contribution is selectively supplied by the few stars hotter than $\sim 10,000$~K. 
In particular, 12 stars may have been detected during their hot HB evolution, in the form
of EHB objects. While nine of them are known targets already classified by \citet{kaluzny92},
three more stars are recognized in our field, being consistent with this scenario (see Paper II
for further discussion).
Coupled with the $(1550-V)$ color-- or its $(FUV-V)$ equivalent-- the H$\beta$ index is found 
to provide a simple and effective diagnostic tool to probe the nature of the UV excess in 
unresolved stellar systems. A ``bluer'' $(1550-V)$ color, as in moderately star-forming ellipticals,
for instance, will always accompany an enhanced H$\beta$ absorption (indicatively, 
$H\beta \ge 2.0$~\AA). The UV-upturn best candidates, therefore, stand out for their relatively 
lower H$\beta$, a definite sign of quiescent evolution.

A comparison with the \citet{dcruz96} theoretical stellar tracks indicates that the hot
edge of the HB stellar distribution in Fig.~\ref{f3} consistently agrees with the evolutionary track 
of a $0.45~M_\odot$ star. Accordingly, most of the \citet{kaluzny92} stars seem currently at the 
limiting mass to ignite Helium in their core \citep{dorman95}, and they may be evolving to the 
WD cooling sequence as AGB-{\it manqu\'e} objects \citep[][see, again, Fig.~\ref{f3}]{greggio90}.
Quite interestingly, our arguments support the claimed evidence of \citet{kalirai07} for a 
prevailing fraction of He-rich low-mass stars with $M\sim 0.43\pm 0.06~M_\odot$ in the WD population 
of NGC~6791. Compared with the expected stellar mass at the Turn Off point 
\citep[about $1.07~M_\odot$, according to][]{grundahl08}, such a reduced WD mass range 
evidently calls for an efficient and pervasive mechanism of mass loss, along the PMS evolution.
While stellar wind may obviously play a role, a combined series of arguments in our analysis
may rather lead to prefer an alternative scenario, where mass transfer mechanisms, 
dealing with binary-system evolution, might likely provide a viable channel to let stellar envelope
vanish along the full PMS evolutionary path \citep{carraro95,carraro96}. One point in this sense 
is certainly the high
($30\pm 10$\%) fraction of binary stars in the cluster \citep{bedin08} and, at the same time,
the lack of any explicit effect of enhanced stellar wind along red-giant evolution
\citep{vanloon08,miglio12}. Rather than calling for any sort of ``superwind'' peeling-off process,
as the analysis of \citet{kalirai07} may for instance lead to envisage, the recognized evidence 
of a reduced lifetime and a correspondingly scanty population of RGB+AGB stars, could more 
naturally be in consequence of a Helium-rich chemical mix for the cluster stellar population. 
By relying on the \citet{iben68} $R^\prime$ parameter, in fact, red-giant number counts are 
consistent with $Y_{\rm 6791} = 0.30${\scriptsize $\pm 0.04$}, where the upper limit of the
allowed range could be preferred, considering the higher statistical probability
for the few bright red giants to be field interlopers.

In this framework, it is interesting to remark the excellent agreement between the implied
lifetime of the different evolutionary branches in the cluster H-R diagram, as derived
from the FCT theorem \citep{rb86} and the theoretical expectations from both the
{\sc Padova} \citep{bertelli08} and {\sc Basti} \citep{pietrinferni04} stellar tracks,
as summarized in Table~\ref{tab:t3}. Both RHB and EHB stars are therefore to be considered 
as a part of the same evolutionary
stage, that physically coincides with the core He-burning phase, and lasts in NGC~6791 a total of
$t_{\rm HB} = 114${\scriptsize $\pm 15$}~Myr. One has to notice, however, that
the energetic arguments, alone, cannot explain the puzzling evidence for 
HB stars to display a so marked mass bimodality. Again, shocking ``outside-in'' physical
mechanisms, possibly related to binary mass transfer, might be addressed to provide
a more satisfactory solution in this sense \citep{yong00,yi04}.
If this is the real case, that applies to UV-enhanced ellipticals too, then one may be led to 
conclude that the UV upturn phenomenon is a quite delicate result of tuned environment conditions
(i.e.\ higher surface brightness and more ``packed'' stars) and evolutionary constraints 
(i.e.\ old and metal-rich stellar populations) inside galaxies. The marginal evidence for 
high-mass ellipticals to better display an UV excess in their SED is a clue in this sense
\citep{burstein88,yi11}.

As the expected timescale for AGB completion
turns about 16-17~Myr, according to the theoretical stellar tracks, then observations
lead to infer for the RGB alone a total lifetime of $t_{\rm RGB} = 649${\scriptsize $\pm 36$}~Myr,
whose 53{\scriptsize $\pm 5$}~Myr are spent at a brighter luminosity than the RHB clump.
As for the MS properties, its reduced photometric contribution (just a 10\% of the total
cluster luminosity, in our sample) seems a firm result, that is not substantially recovered
even by accounting for lost faint stars, and rather points to an inherently dwarf-depleted
IMF, consistent or flatter than the Salpeter case.

In force of the FCT, by summing up the photometric contribution of stars along the different 
PMS phases, we derive from Table~\ref{tab:t3} a total fuel consumption of   
0.43{\scriptsize$\pm 0.01$}~M$_\odot$. This figure is in close agreement with the expected
He core from theoretical tracks, and with the estimated WD mass of  
$\langle m_{\rm WD}\rangle = 0.43 \pm 0.06$~M$_\odot$, indicating that
cluster stars fully exhaust (or lose) their external envelope along the full PMS
evolution. Such a tight agreement, may also lead to suspect, with \citet{kalirai07}, that a 
fraction of the cluster stellar population does not reach the minimum mass required for
stars to effectively ignite He in their cores.\\

\acknowledgments

One of us (A.B.) would like to thank the Instituto Nacional de Astrofisica Optica
y Electronica of Puebla (Mexico) and the European Southern Observatory for
generous financial support and a warm hospitality during several visits in Mexico
and at the ESO premises in Santiago de Chile, where part of this work has been 
conceived. Partial financial support by Mexican SEP-CONACyT grant 47904
is also acknowledged.

This work has made use of different on-line galactic and extragalactic 
databases, including the {\sc Webda} database for stellar clusters in the Galaxy and the 
Magellanic Clouds, maintained at the Institute of Astronomy of the University 
of Vienna, the Hyper-Linked Extragalactic Databases and Archives (HyperLeda) based 
at the Lyon University, and the VizieR catalog service of the Centre de Donn\'ees 
astronomiques de Strasbourg.

\clearpage
\end{document}